\pdfoutput=1

\documentclass[11pt,twoside,a4paper,cmspaper,final,collab]{cms-tdr}
\def\svnVersion{485963P}\def\cmsCernNoTag{CERN-EP-2018-251}\def\cmsCernDate{\today}\def\cmsMessage{Published in Physical Review D as \href{http://dx.doi.org/10.1103/PhysRevD.99.012005}{\doi{10.1103/PhysRevD.99.012005.}}}

\begin{document}\cmsNoteHeader{EXO-17-024}

\hyphenation{had-ron-i-za-tion}
\hyphenation{cal-or-i-me-ter}
\hyphenation{de-vices}
\RCS$HeadURL: svn+ssh://svn.cern.ch/reps/tdr2/papers/EXO-17-024/trunk/EXO-17-024.tex $
\RCS$Id: EXO-17-024.tex 483734 2018-12-06 15:28:36Z woodson $
\newlength\cmsSmallFigWidth
\newlength\cmsFigWidth
\newlength\cmsTabWidth
\newlength\cmsTabSkip\setlength{\cmsTabSkip}{1ex}
\ifthenelse{\boolean{cms@external}}{\setlength\cmsSmallFigWidth{0.39\textwidth}}{\setlength\cmsSmallFigWidth{0.43\textwidth}}
\ifthenelse{\boolean{cms@external}}{\setlength\cmsFigWidth{0.45\textwidth}}{\setlength\cmsFigWidth{0.49\textwidth}}
\ifthenelse{\boolean{cms@external}}{\setlength\cmsTabWidth{\textwidth}}{\setlength\cmsTabWidth{\textwidth}}
\ifthenelse{\boolean{cms@external}}{\providecommand{\cmsLeft}{upper\xspace}}{\providecommand{\cmsLeft}{left\xspace}}
\ifthenelse{\boolean{cms@external}}{\providecommand{\cmsRight}{lower\xspace}}{\providecommand{\cmsRight}{right\xspace}}
\ifthenelse{\boolean{cms@external}}{\providecommand{\CL}{C.L.\xspace}}{\providecommand{\CL}{CL\xspace}}
\ifthenelse{\boolean{cms@external}}{\providecommand{\NA}{\ensuremath{\cdots}\xspace}}{\providecommand{\NA}{\ensuremath{\text{---}}\xspace}}
\newcommand{\mSD}{\ensuremath{m_{\mathrm{SD}}}\xspace}
\newcommand{\Rpf}{\ensuremath{R_{\mathrm{p}/\mathrm{f}}}\xspace}
\newcommand{\nsub}{\ensuremath{\tau_{21}}\xspace}
\newcommand{\nsubddt}{\ensuremath{\tau_{21}^{\mathrm{DDT}}}\xspace}
\newcommand{\nddt}{\ensuremath{\mathrm{N}_{\mathrm{2}}^{\mathrm{1,DDT}}}\xspace}
\newcommand{\gqphi}{\ensuremath{g_{\Pq\Phi}}\xspace}
\newcommand{\PA}{\ensuremath{\cmsSymbolFace{A}}}
\newcommand{\PV}{\ensuremath{\cmsSymbolFace{V}}}
\newcommand{\gqa}{\ensuremath{g_{\Pq\PA}}\xspace}
\newcommand{\cPf}{\ensuremath{\cmsSymbolFace{f}}}
\newcommand{\cPaf}{\ensuremath{\cmsSymbolFace{\overline{f}}}}

\cmsNoteHeader{EXO-17-024}
\title{Search for low-mass resonances decaying into bottom quark-antiquark pairs in proton-proton collisions at \texorpdfstring{$\sqrt{s} = 13\TeV$}{sqrt(s) = 13 TeV}}
\date{\today}

\abstract{
A search for narrow, low-mass, scalar and pseudoscalar resonances decaying to bottom quark-antiquark pairs is presented. The search is based on events recorded in $\sqrt{s}=13\TeV$ proton-proton collisions with the CMS detector at the LHC, collected in 2016, and corresponding to an integrated luminosity of 35.9\fbinv. The search selects events in which the resonance would be produced with high transverse momentum because of the presence of initial- or final-state radiation. In such events, the decay products of the resonance would be reconstructed as a single large-radius jet with high mass and two-prong substructure. A potential signal would be identified as a narrow excess in the jet invariant mass spectrum. No evidence for such a resonance is observed within the mass range from 50 to 350\GeV, and upper limits at 95\% confidence level are set on the product of the cross section and branching fraction to a bottom quark-antiquark pair. These constitute the first constraints from the LHC on exotic bottom quark-antiquark resonances with masses below 325\GeV.
}

\hypersetup{
pdfauthor={CMS Collaboration},
pdftitle={Search for low-mass resonances decaying into bottom quark-antiquark pairs in proton-proton collisions at sqrt(s) = 13 TeV},
pdfsubject={CMS},
pdfkeywords={CMS, boosted dijet, scalar, pseudoscalar, bottom quark-antiquark decay, Higgs, double-b tagger}}

\maketitle
\section{Introduction}
\label{sec:intro}

Many models of physics beyond the standard model (SM) require new particles that couple to quarks and gluons and can be observed as dijet resonances. One example is a model in which dark matter particles ($\chi$) couple to SM particles through a spin-0 scalar ($\Phi$) or pseudoscalar ($\PA$) mediator, which decays preferentially to a bottom quark-antiquark ($\bbbar$) pair~\cite{Abercrombie:2015wmb,Buckley:2014fba,Harris:2014hga,Haisch:2015ioa,Boveia:2016mrp}.
As the mass of such a mediator is an unknown parameter of the model, it is important to search in as broad a mass range as possible.

Because of the overwhelming background of events from jets produced through the strong interaction, referred to as quantum chromodynamics (QCD) multijet events, inclusive searches for dijet resonances at the CERN LHC have historically been limited to dijet invariant masses greater than 1\TeV. Several techniques have been explored to evade this limitation. Trigger-level analyses, also known as ``data scouting,'' increase the number of events collected at lower dijet invariant masses by recording a minimal subset of the total event content.
The ATLAS and CMS experiments have used this technique to search for resonances with masses as low as 450\GeV~\cite{CMS11,Sirunyan:2016iap,Sirunyan:2018xlo,Aaboud:2018fzt}. The invariant mass threshold can also be lowered by performing bottom quark tagging at the trigger level, enabling masses as low as 325\GeV to be probed~\cite{Sirunyan:2018pas,Aaboud:2018tqo}.
The analysis presented here uses a different technique, requiring that the dijet resonances be produced with significant initial- or final-state radiation. The technique has been employed in searches for low mass resonances decaying to quark-antiquark pairs~\cite{Sirunyan:2017dnz,Sirunyan:2017nvi,Aaboud:2018zba}, which have provided the best sensitivity to date for resonances with masses between 50 and 300\GeV. This technique has also been used to search for SM Higgs bosons ($\PH$) produced through gluon fusion and decaying to $\bbbar$ pairs~\cite{Sirunyan:2017dgc}, with an observed significance of 1.5 standard deviations.

This paper presents the first LHC search for new particles that decay to $\bbbar$ resonances with masses as low as 50\GeV.
Spin-0 scalar and pseudoscalar resonances, which may mediate interactions between dark matter particles and SM particles, are considered.
Minimal flavor violation is assumed, to ensure consistency with flavor constraints~\cite{Abercrombie:2015wmb,Buckley:2014fba,Harris:2014hga,Haisch:2015ioa,Boveia:2016mrp}.
Under this assumption, the $\Phi$ or $\PA$ particles decay only to fermion-antifermion pairs of the same flavor.
Further, the SM couplings are assumed to be proportional to the SM Yukawa couplings with a single universal constant of proportionality, $\gqphi$ or $\gqa$.
The two interaction Lagrangians are
\begin{align}
\mathcal{L}_{\Phi} &= g_{\chi\Phi} \Phi \overline{\chi} \chi  +
	\frac{\Phi}{\sqrt{2}} \sum_{\cPf} \gqphi y_{\cPf} \cPaf \cPf , \label{eq:scalarlag} \\
\mathcal{L}_{\PA} & =  ig_{\chi\PA} \PA \overline{\chi} \gamma_5\chi +
	\frac{i \PA}{\sqrt{2}}\sum_{\cPf}  \gqa y_{\cPf} \cPaf \gamma_5 \cPf , \label{eq:pseudoscalarlag}
  \end{align}
where the sum is over all charged SM fermions, $g_{\chi\Phi}$ and $g_{\chi\PA}$ are the couplings to the dark matter particle, the Yukawa couplings of fermions $y_{\cPf}$ are normalized to the Higgs vacuum expectation value as $y_{\cPf} = \sqrt{2}m_{\cPf}/v$ with $v=246\GeV$ and $m_{\cPf}$ the corresponding fermion mass.
For resonance masses below twice the dark matter particle mass ($m_\chi$), $\Phi$ and $\PA$ couple preferentially to heavier quarks.
Consequently, the resonances are predominantly produced via a loop-induced coupling to gluons, and, for resonance masses below twice the top quark mass ($m_\cPqt$), decay mostly to $\bbbar$ pairs.
This search is also sensitive to extensions of the SM that include a new gauge boson that couples to the right-handed components of the bottom and charm quarks~\cite{Liu:2017xmc}.

This paper reports the results of a search for narrow $\bbbar$ resonances with masses between 50 and 350\GeV in events collected in $\sqrt{s}=13\TeV$ proton-proton ($\Pp\Pp$) collisions with the CMS detector at the LHC.
The data sample corresponds to an integrated luminosity of $35.9\fbinv$.
We search for resonances produced with high transverse momentum $\pt$ because of significant initial- or final-state radiation (ISR or FSR).
This ISR or FSR ensures the events pass stringent trigger restrictions set by bandwidth limitations, allowing resonance masses as low as 50\GeV to be probed.
The resonance decay products are merged into a single wide jet.
Two wide-jet algorithms are considered: the anti-\kt algorithm~\cite{Cacciari:2008gp} with distance parameter $R = 0.8$ (AK8), and the Cambridge--Aachen algorithm~\cite{Dokshitzer:1997in,Wobisch:1998wt} with distance parameter $R = 1.5$ (CA15).
The AK8 algorithm provides better sensitivity at signal masses below 175\GeV, while the CA15 algorithm provides better sensitivity at higher masses because of the increased acceptance of decay products with wider angular separation~\cite{Butterworth:2008iy}.
Jet substructure~\cite{Moult:2016cvt} techniques and dedicated $\cPqb$ tagging~\cite{Sirunyan:2017ezt} algorithms are used to distinguish the signal from the QCD background.

\section{The CMS detector}
The central feature of the CMS apparatus is a superconducting solenoid of 6\unit{m} internal diameter, providing a magnetic field of 3.8\unit{T}. A silicon pixel and strip tracker, a lead tungstate crystal electromagnetic
calorimeter, and a brass and scintillator hadron calorimeter, each composed of a barrel and two endcap sections, reside within the solenoid. Forward calorimeters extend the pseudorapidity ($\eta$) coverage provided by the barrel and endcap detectors. Muons are detected in gas-ionization chambers embedded in the steel flux-return yoke outside the solenoid.

Events of interest are selected using a two-tiered trigger system~\cite{Khachatryan:2016bia}. The first level, composed of custom hardware processors, uses information from the calorimeters and muon detectors to select events at a rate of around 100\unit{kHz} within a time interval of less than 4\mus. The second level, known as the high-level trigger, consists of a farm of processors running a version of the full event reconstruction software optimized for fast processing, and reduces the event rate to around 1\unit{kHz} before data storage.

A more detailed description of the CMS
detector, together with a definition of the coordinate system used and
the relevant kinematic variables, can be found
in Ref.~\cite{Chatrchyan:2008zzk}.
\label{sec:detector}

\section{Simulated samples}
\begin{figure}[htbp]
\centering
\includegraphics[width=\cmsFigWidth]{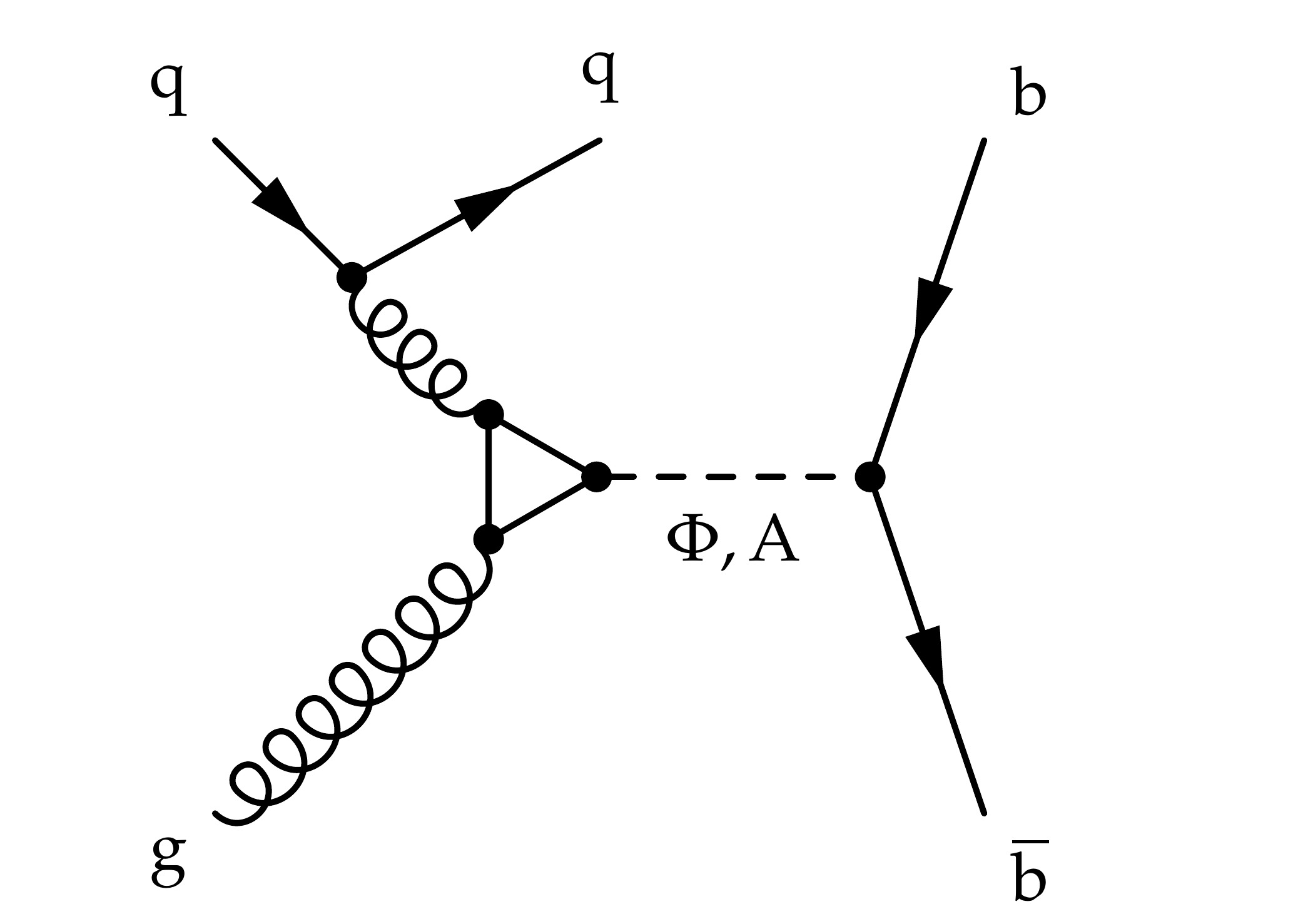}
\includegraphics[width=\cmsFigWidth]{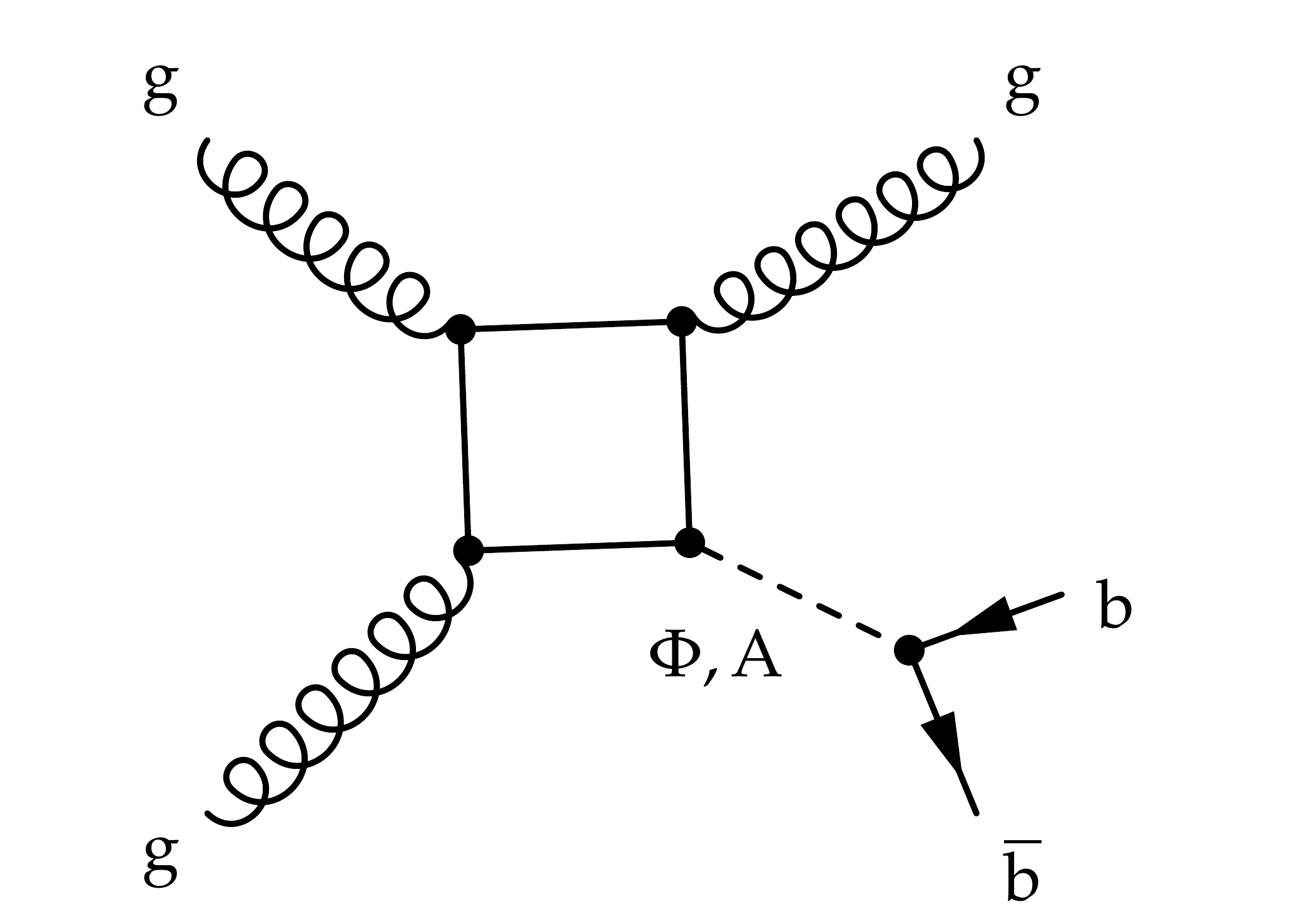}
\caption{One-loop Feynman diagrams of processes exchanging a scalar $\Phi$ (\cmsLeft) or pseudoscalar $\PA$ (\cmsRight) mediator, leading to a boosted double-$\cPqb$ jet signature.}
\label{fig:feyn_prod_S}
\end{figure}

Simulated samples of signal and background events are produced using various Monte Carlo (MC) event generators, with the CMS detector response modeled by \GEANTfour~\cite{GEANT4}.
The benchmark $\Phi$ and $\PA$ signal events, produced primarily via gluon fusion, are simulated using the {\MGvATNLO}~2.4.2 generator~\cite{Alwall:2014hca} for various mass hypotheses in the range 50--500\GeV. The events are generated with a parton-level filter requiring total hadronic transverse energy $\HT>400\GeV$; events failing this requirement fall outside the acceptance of the analysis selection, discussed in the following section.
Figure~\ref{fig:feyn_prod_S} shows representative one-loop Feynman diagrams producing a boosted jet originating from a $\bbbar$ pair (double-$\cPqb$ jet).

\label{sec:samples}

In accordance with the recommendations of the ATLAS-CMS Dark Matter Forum~\cite{Abercrombie:2015wmb} and the LHC Dark Matter Working Group~\cite{Boveia:2016mrp}, the $\Phi$ and $\PA$ signal samples are normalized to their production cross sections at leading order (LO) accuracy calculated with the {\MGvATNLO} 2.4.2 generator using the DMSimp package~\cite{Mattelaer:2015haa}.
The total cross sections, which are compared to the upper limits obtained with this analysis, are calculated using the LO diagram with no additional partons and no cuts applied to the final state kinematics.
The production cross section at next-to-leading order (NLO) accuracy including the finite $m_\cPqt$ has only been calculated for a scalar with a mass of 125\GeV, where it is approximately a factor of 2 greater~\cite{Jones:2018hbb}.
This NLO correction is not used in this analysis; applying it would not affect the sensitivity of the search to the signal production cross section, but it would improve the sensitivity to the couplings $\gqphi$ or $\gqa$ by a factor of approximately $\sqrt{2}$.

The \MGvATNLO~2.3.3~\cite{Alwall:2014hca} generator is used for the diboson, $\PW+$jets, $\cPZ+$jets, and QCD multijet samples, at LO accuracy with matching~\cite{Alwall:2007fs} between jets from the matrix element calculation and the parton shower description,
while \POWHEG 2.0~\cite{Nason:2004rx,Frixione:2007vw,Alioli:2010xd} at NLO precision is used to model the $\ttbar$ and single top processes.
The Higgs boson signal samples are produced using the {\POWHEG}+ {\sc MiNLO}~\cite{Frixione:2007vw,Luisoni:2013kna} event generator with $m_{\PH} = 125\GeV$.
For the gluon fusion production mode, the \POWHEG generated sample with up to one extra jet in matrix element calculations is normalized to the inclusive cross section at next-to-next-to-next-to-leading order (N$^{3}$LO) accuracy~\cite{deFlorian:2012mx,Grazzini:2013mca,Bagnaschi2012,Bagnaschi:2015qta}, with a $\pt$-dependent correction to account for the effects of the finite $m_\cPqt$ and associated higher-order QCD corrections~\cite{Sirunyan:2017dgc}.

For parton showering and hadronization, the \POWHEG and \MGvATNLO samples are interfaced with \PYTHIA 8.212~\cite{Sjostrand:2014zea}.
The \PYTHIA parameters for the underlying event description are set to the
CUETP8M1 tune~\cite{Khachatryan:2015pea}.
The production cross sections for the diboson samples are calculated to next-to-next-to-leading order (NNLO) accuracy with the \textsc{mcfm}~7.0 program~\cite{Campbell:2010ff}. The cross section for top quark pair production is computed with \textsc{Top++} 2.0~\cite{Czakon:2013goa} at NNLO including soft-gluon resummation to next-to-next-to-leading-log order. The cross sections for $\PW+$jets and $\cPZ+$jets samples include higher-order QCD and electroweak (EW)
corrections improving the modeling of high-$\pt$ $\PW$ and $\PZ$ bosons events~\cite{Kallweit:2014xda,Kallweit:2015dum,Kallweit:2015fta,Lindert:2017olm}.
The parton distribution function set {NNPDF3.0}~\cite{Ball:2014uwa} is used to produce all simulated samples, with the accuracy (LO or NLO) corresponding to that of the matrix elements used for generation.

\section{Event reconstruction and selection}
\label{sec:physobj}

Event reconstruction is based on a particle-flow algorithm~\cite{Sirunyan:2017ulk}, which aims to reconstruct and identify each individual particle with an optimized combination of information from the various elements of the CMS detector.
The algorithm identifies each reconstructed particle as an electron, a muon, a photon, or a charged or neutral hadron. The missing transverse momentum vector is defined as the negative vector sum of the transverse momenta of all the particles identified in the event, and its magnitude is referred to as $\ptmiss$. Particles are clustered into AK8~\cite{Cacciari:2008gp} or CA15~\cite{Dokshitzer:1997in} jets, depending on the signal mass hypothesis.
The clustering algorithms are implemented by the \textsc{FastJet} package~\cite{Cacciari:2011ma}.
To mitigate the effect from the contributions of extraneous $\Pp\Pp$ collisions (pileup), the pileup per particle identification algorithm~\cite{Bertolini:2014bba} assigns a weight to each particle prior to jet clustering based on the likelihood of the particle to originate from the hard scattering vertex.
Further corrections are applied to the jet energy as a function of jet $\eta$ and $\pt$  to bring the measured response of jets to that of particle level jets on average~\cite{Khachatryan:2016kdb}.

A combination of several event selection criteria is used to trigger on events, all imposing minimum thresholds either on $\HT$ or on the AK8 jet $\pt$. In addition, a minimum threshold on the trimmed jet mass, where remnants of soft radiation are removed before computing the mass~\cite{Krohn:2009th}, is imposed to reduce the $\HT$ or $\pt$ thresholds and improve the signal acceptance.
The trigger selection is greater than 95\% efficient at selecting events with at least one AK8 jet with $\pt > 450\GeV$, $\abs{\eta} < 2.5$, and mass greater than 40\GeV or events with at least one CA15 jet with $\pt > 500\GeV$ and $\abs{\eta} < 2.5$.
We also define six (five) $\pt$ categories from 450 (500)\GeV to 1\TeV for AK8 (CA15) jets with variable width from 50 to 200\GeV.
To reduce backgrounds from SM EW processes, events containing isolated electrons~\cite{Khachatryan:2015hwa} or muons~\cite{Sirunyan:2018fpa}, or hadronically decaying $\tau$ leptons with $\pt > 10$, 10, or 18\GeV and $\abs{\eta} < 2.5$, 2.4, or 2.3, respectively, are vetoed. For electrons or muons, the isolation criteria require that the pileup-corrected sum of the $\pt$ of charged hadrons and neutral particles surrounding the lepton
divided by the lepton $\pt$ be less than approximately 15 or 25\%, respectively, depending on $\eta$~\cite{Khachatryan:2015hwa,Sirunyan:2018fpa}.
Events with $\ptmiss >140\GeV$ are vetoed in order to reduce the top quark background contamination.
For each event, the leading jet in $\pt$ is assumed to be the $\Phi(\bbbar)$ or $\PA(\bbbar)$ candidate.
The soft-drop algorithm~\cite{Larkoski:2014wba} with angular exponent $\beta=0$ is applied to the jet to remove soft and wide-angle radiation with a soft radiation fraction $z$ less than 0.1.
The parameter $\beta$ controls the grooming profile as a function of subjet separation; when $\beta=0$, the grooming threshold is independent of subjet separation, and the algorithm is equivalent to the modified mass-drop tagger~\cite{Dasgupta:2013ihk}.
For background QCD multijet events where large jet masses arise from soft gluon radiation, the soft-drop jet mass $\mSD$ is reduced relative to the ungroomed jet mass.
On the other hand, for signal events $\mSD$ is primarily determined by the $\Phi(\bbbar)$ decay kinematics, and the distribution peaks at the mass of the $\Phi(\bbbar)$ signal.

Dedicated $\mSD$ corrections are derived from a comparison of simulated and measured samples in a region enriched with merged
$\PW(\qqbar)$ decays from $\ttbar$ events~\cite{CMS-PAS-JME-16-003}. The $\mSD$ corrections remove a residual dependence on the jet $\pt$, and match the simulated jet mass scale and resolution to those observed in data.
A lower $\mSD$ bound of 40\GeV is applied in the search with AK8 jets to ensure that the trigger has greater than 95\% efficiency, while a lower $\mSD$ bound of 82\GeV is applied in the search with CA15 jets to ensure the background model described in Section~\ref{sec:bkgest} is robust.
The resulting $\mSD$ distributions are binned with a bin width of 7\GeV, corresponding to the $\mSD$ resolution near the $\PW$ and $\PZ$ resonances.

The dimensionless mass scale variable for QCD multijet jets, $\rho=\ln(\mSD^2/\pt^2)$~\cite{Dasgupta:2013ihk,Dolen:2016kst}, is used to characterize the correlation between the jet $\cPqb$ tagging discriminator, jet mass, and jet $\pt$. Its distribution is roughly invariant in different ranges of jet $\pt$.
Only events in the range $-6.0 < \rho < -2.1$ ($-4.7 < \rho < -1.0$) are considered for AK8 (CA15) jets, effectively defining different $\mSD$ ranges depending on jet $\pt$.
The upper bound is imposed to avoid instabilities at the edges of the distribution due to finite cone limitations from the jet clustering, while the lower bound avoids the nonperturbative regime of the $\mSD$ calculation.
This requirement is about 98\% efficient for the $\Phi(\bbbar)$ signal at low masses (50--125\GeV) when reconstructed as an AK8 jet, and 60--85\% efficient at high masses (200--350\GeV) when reconstructed as a CA15 jet.

The $N_2^{1}$ variable~\cite{Moult:2016cvt} is used to determine how consistent a jet is with having a two-prong substructure.
It is based on a ratio of 2-point ($_{1}e_{2}$) and 3-point  ($_{2}e_{3}$) generalized energy correlation functions~\cite{Larkoski:2013eya},
\ifthenelse{\boolean{cms@external}}{
\begin{align}
_{1}e_{2} &= \sum_{1\leq i < j \leq n} z_{i}z_{j}\Delta R_{ij} , \\
_{2}e_{3} &= \sum_{1\leq i < j < k\leq n} z_{i}z_{j}z_{k} \nonumber \\
&\times \min \{\Delta R_{ij}\Delta R_{ik}, \Delta R_{ij}\Delta R_{jk}, \Delta R_{ik}\Delta R_{jk} \} ,
\end{align}
}
{
\begin{align}
_{1}e_{2} &= \sum_{1\leq i < j \leq n} z_{i}z_{j}\Delta R_{ij} , \\
_{2}e_{3} &= \sum_{1\leq i < j < k\leq n} z_{i}z_{j}z_{k} \min \{\Delta R_{ij}\Delta R_{ik}, \Delta R_{ij}\Delta R_{jk}, \Delta R_{ik}\Delta R_{jk} \} ,
\end{align}
}
where $z_i$ represents the energy fraction of the constituent $i$ in the jet and $\Delta R_{ij}$
is the angular separation between constituents $i$ and $j$.
These generalized energy correlation functions ${_{v}e_n}$ are sensitive to correlations of $v$ pairwise angles among $n$-jet constituents~\cite{Moult:2016cvt}.
For a two-prong structure, signal jets have a stronger 2-point correlation than a 3-point correlation.
The discriminant variable $N_2^{1}$ is then constructed via the ratio:
\begin{equation}
\quad N_{2}^{1}= \frac{_{2}e_{3}}{(_{1}e_{2})^{2}} .
\end{equation}

The calculation of $N_2^{1}$ is based on the jet constituents after application of the soft-drop grooming algorithm to the jet.
It provides excellent discrimination between two-prong signal jets and QCD background jets.
However, imposing requirements on $N_2^{1}$, or other similar variables, distorts the jet mass distributions differently depending on the $\pt$ of the jet~\cite{Thaler:2010tr}.
To minimize this distortion, a transformation is applied to $N_{2}^{1}$ following the designed decorrelated tagger (DDT) technique~\cite{Dolen:2016kst} to reduce its correlation with $\rho$ and $\pt$ in multijet events.
The transformed variable is defined as $\nddt \equiv N_2^{1} - X_{(26\%)}$, where $X_{(26\%)}$ is the 26th percentile of the $N_2^{1}$ distribution in simulated QCD events as a function of $\rho$ and $\pt$.
The transformation is derived in bins of $\rho$ and $\pt$, separately for AK8 and CA15 jets.
This ensures that the selection $\nddt<0$ yields a constant QCD background efficiency across the $\rho$ and $\pt$ range considered in this search.
The chosen background efficiency of 26\% maximizes the signal sensitivity, independent of the signal mass.

A dedicated double-$\cPqb$ tagger is used to select jets likely to originate from two $\cPqb$ quarks~\cite{Sirunyan:2017ezt}.
Events where the selected wide jet is double-$\cPqb$-tagged constitute the ``passing,'' or signal, region, while events failing the double-$\cPqb$ tagger form the ``failing'' region, which is used to estimate the QCD multijet background in the signal region.
The multivariate algorithm, based on a boosted decision tree, takes as inputs several observables that characterize the distinct properties of $\cPqb$ hadrons and their flight directions in relation to the jet substructure. A wide jet is considered double-$\cPqb$ tagged if its double-$\cPqb$ tagger discriminator value exceeds a threshold corresponding to a 1\% misidentification rate for QCD jets and a 33\% efficiency for $\Phi(\bbbar)$ candidates with a mass of 125\GeV reconstructed as AK8 jets.

For CA15 jets, because of the larger cone with radius parameter of 1.5, it is often possible to resolve two subjets within the wide jet; hence additional background discrimination can be obtained by incorporating the individual subjet $\PQb$ tagging probabilities. The subjets are constructed using the soft-drop algorithm, and assigned $\cPqb$ tagging scores using the combined secondary vertex algorithm (CSVv2)~\cite{Sirunyan:2017ezt} that combines information from displaced tracks and vertices using a multilayer perceptron. The second highest CSVv2 score is then used as an additional input to the boosted decision tree of the double-$\cPqb$ tagger.
For the chosen discriminator threshold, the double-$\cPqb$ tagger algorithm has a misidentification rate of about 4\%, and a signal efficiency which decreases with mass, equalling 25 (13)\% for a signal mass of 200\GeV (350\GeV).

The efficiency (in percent) of the cumulative selection criteria for the scalar $\Phi(\bbbar)$ signal benchmark is shown in Table~\ref{tab:CutFlow}. The efficiencies for the $\PA(\bbbar)$ signal are consistent within the statistical uncertainties.

\begin{table*}
	\centering
	\topcaption{The selection efficiencies in percent for simulated $\Phi(\bbbar)$ signal events with parton-level $\HT>400\GeV$, at different stages of the event selection, shown for different mass hypotheses and for AK8 and CA15 jets. The statistical uncertainties due to the simulated sample size are also shown.}
	\label{tab:CutFlow}
	\resizebox{\cmsTabWidth}{!}{
		\begin{scotch}{c D{,}{\,\pm\,}{3.2} D{,}{\,\pm\,}{3.2} D{,}{\,\pm\,}{3.2} D{,}{\,\pm\,}{3.2} D{,}{\,\pm\,}{3.2} D{,}{\,\pm\,}{3.2} D{,}{\,\pm\,}{2.2}}
			\multicolumn{8}{c}{AK8 jets} \\
			 \multirow{2}{*}{$m_{\Phi}$ (GeV)}	&	\multicolumn{1}{c}{$\pt$}	&	\multicolumn{1}{c}{$\mSD$}	&	\multicolumn{1}{c}{\multirow{2}{*}{Lepton veto}}	&	\multicolumn{1}{c}{$\ptmiss$}	&	\multicolumn{1}{c}{\multirow{2}{*}{$\nddt<0$}} & \multicolumn{1}{c}{\multirow{2}{*}{$-6<\rho<2.1$}} & \multicolumn{1}{c}{\multirow{2}{*}{double-$\cPqb$ tag}}  \\
			 	&	\multicolumn{1}{c}{$>450\GeV$}	&	\multicolumn{1}{c}{$>40\GeV$}	&	&	\multicolumn{1}{c}{$<140\GeV$}	&	 &  &  \vspace{\cmsTabSkip}\\
			50	&	75.0 , 0.1	&	37.5 , 0.2	&	36.2 , 0.2	&	32.9 , 0.2	&	14.7 , 0.1	&	14.3 , 0.1	&	7.3 , 0.1	\\
			100	&	75.4 , 0.1	&	42.2 , 0.2	&	40.6 , 0.2	&	37.5 , 0.2	&	18.0 , 0.1	&	17.5 , 0.1	&	7.1 , 0.1	\\
			125	&	75.5 , 0.2	&	42.3 , 0.2	&	40.6 , 0.2	&	37.5 , 0.2	&	18.1 , 0.1	&	17.5 , 0.1	&	6.1 , 0.1	 \vspace{\cmsTabSkip}\\
			\noalign{\vskip \cmsTabSkip}
			\multicolumn{8}{c}{CA15 jets} \\
			 \multirow{2}{*}{$m_{\Phi}$ (GeV)}	&	\multicolumn{1}{c}{$\pt$}	&	\multicolumn{1}{c}{$\mSD$}		&	\multicolumn{1}{c}{\multirow{2}{*}{Lepton veto}}	&	\multicolumn{1}{c}{$\ptmiss$}	&	\multicolumn{1}{c}{\multirow{2}{*}{$\nddt<0$}}	&	\multicolumn{1}{c}{\multirow{2}{*}{$-4.7<\rho<-1.0$}}	&	\multicolumn{1}{c}{\multirow{2}{*}{double-$\cPqb$ tag}} \\
			 	&	\multicolumn{1}{c}{$>500\GeV$}	&	\multicolumn{1}{c}{$>82\GeV$}		&	&	\multicolumn{1}{c}{$<140\GeV$}	&		&		&	 \vspace{\cmsTabSkip}\\
			200		&	61.0 , 0.1	&	35.6 , 0.1	&	33.9 , 0.1	&	31.1 , 0.1	&	13.9 , 0.1	&	13.0 , 0.1	&	3.3 , 0.1	\\
			300		&	63.4 , 0.1	&	35.7 , 0.1	&	34.0 , 0.1	&	31.1 , 0.1	&	13.2 , 0.1	&	11.1 , 0.1	&	1.9 , 0.1	\\
			350		&	64.3 , 0.1	&	35.8 , 0.1	&	33.9 , 0.1	&	31.1 , 0.1	&	13.0 , 0.1	&	8.6 , 0.1	&	1.1 , 0.1	\\
		\end{scotch}}
\end{table*}

\section{Background estimation}
\label{sec:bkgest}

The $\PW$, $\PZ$, and $\PH+$jets backgrounds are modeled using MC simulation. Their overall contribution is less than 5\% of the total SM background.
The normalization and shape of the simulated $\PW/\PZ+$jets backgrounds are corrected for NLO QCD and EW effects.
Other EW processes, such as diboson, triboson, and $\ttbar+\PW/\PZ$, are estimated from simulation and found to be negligible.

The contribution of $\ttbar$ production to the total SM background, estimated to be less than 3\%, is obtained from simulation corrected with scale factors derived from a $\ttbar$-enriched control sample in which an isolated muon~\cite{Sirunyan:2018fpa} is required.
Scale factors correct the overall $\ttbar$ normalization and the double-$\cPqb$ mistag efficiency for jets originating from top quark decays.
The control sample is included in the global fit used to extract the signal, with the scale factors treated as unconstrained parameters.

The main background in the passing region, QCD multijet production, has a nontrivial jet mass shape that is difficult to model parametrically and depends on jet $\pt$.
Therefore, we constrain it using the background-enriched failing region, i.e., events failing the double-$\cPqb$ tagger selection.
Since the double-$\cPqb$ tagger discriminator and the jet mass are largely uncorrelated, the passing and failing regions have similar QCD jet mass distributions, and their ratio, the ``pass-fail ratio'' $\Rpf$, is expected to be nearly constant as a function of jet mass and $\pt$.
To account for the residual difference between the shapes of passing and failing events, $\Rpf$ is parametrized as a Bernstein polynomial in $\rho$ and $\pt$,
\begin{align}
\Rpf(\rho,\pt) &= \sum^{n_{\rho}}_{k=0} \sum^{n_{\pt}}_{\ell=0} a_{k, \ell} b_{k, n_{\rho}}(\rho)  b_{\ell, n_{\pt}}(\pt),~
\end{align}
where $n_{\rho}$ is the degree of the polynomial in $\rho$, $n_{\pt}$ is the degree of the polynomial in $\pt$, $a_{k, \ell}$ is a Bernstein coefficient, and
\begin{align}
b_{{\nu ,n}}(x)&=\binom{n}{\nu} x^{{\nu }}\left(1-x\right)^{{n-\nu}}
\end{align}
is a Bernstein basis polynomial of degree $n$.

\begin{figure*}[hbtp]
  \centering
  \includegraphics[width=\cmsFigWidth]{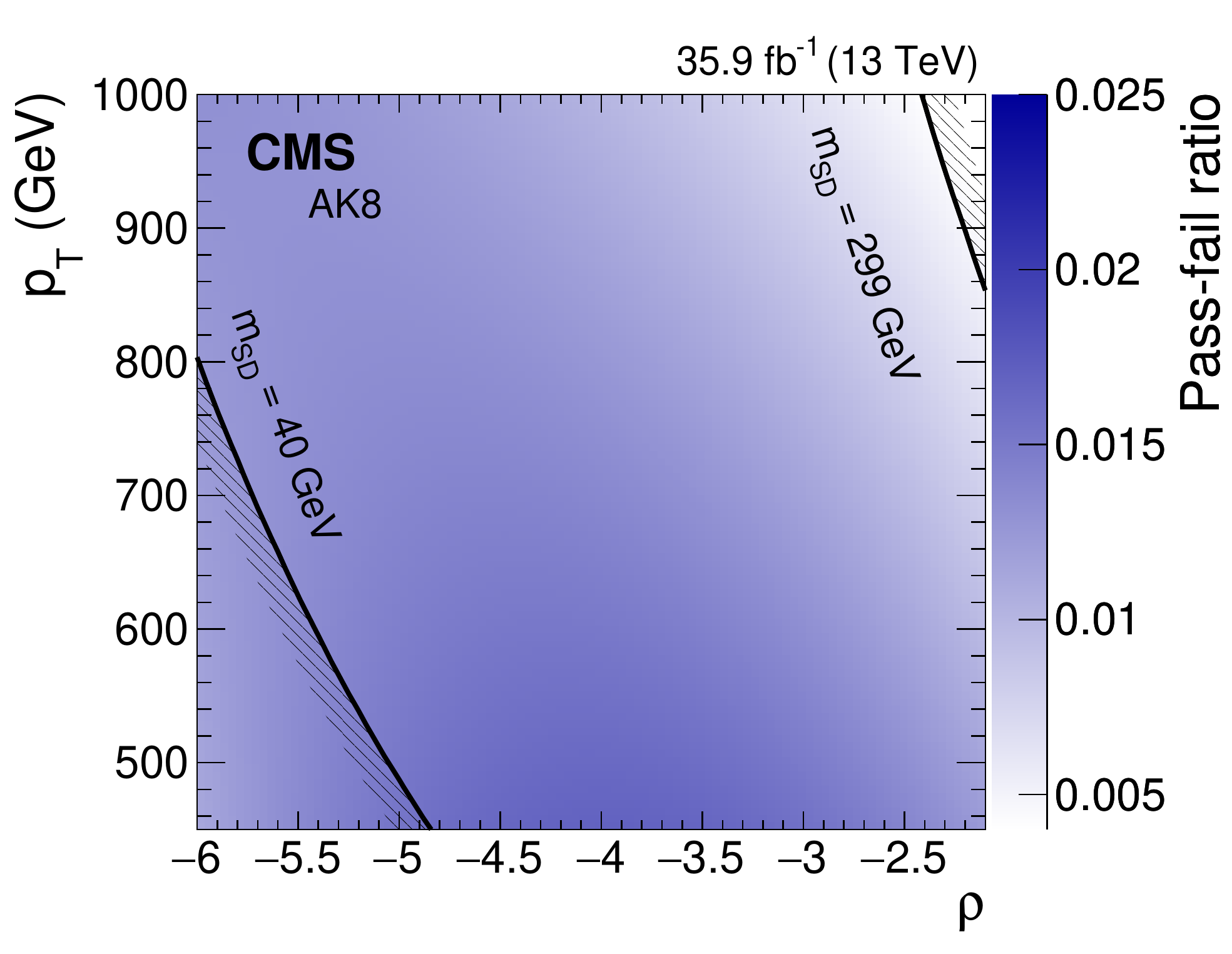}
  \includegraphics[width=\cmsFigWidth]{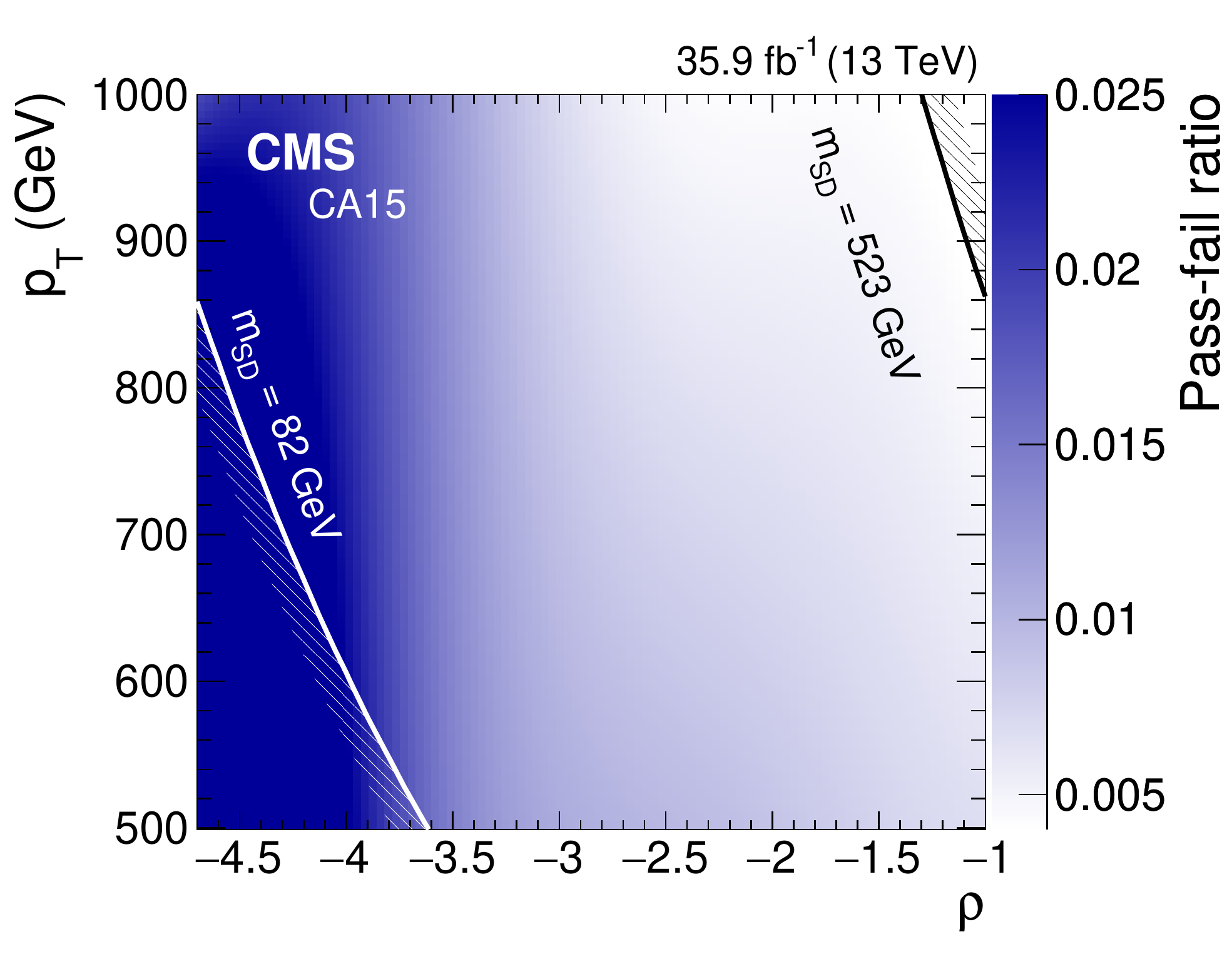}
  \caption{The fitted pass-fail ratio $\Rpf$ as a function of $\pt$ and $\rho$ for the AK8
    selection (left) and the CA15 selection (right).}
  \label{fig:qcdRatio2d}
\end{figure*}

The coefficients $a_{k, \ell}$ have no external constraints, but are determined from a simultaneous binned fit to data in passing and failing regions across the whole jet mass and $\pt$ range.
The $\pt$ binning, varying from 50 to 200\GeV, is chosen to provide enough data points to constrain the shape of $\Rpf$.
To determine the degree of polynomial necessary to fit the data, a Fisher $F$-test~\cite{ref:ftest} is performed.
Based on its results, a polynomial of second (fifth) degree in $\rho$ and first degree in $\pt$ is selected for the AK8 (CA15) analysis category. The fitted pass-fail ratios $\Rpf$ as functions of $\rho$ and $\pt$ under the background-only hypothesis are shown in Fig.~\ref{fig:qcdRatio2d} for the AK8 and CA15 selections.

\begin{figure*}[hbtp]
\centering
    \includegraphics[width=\cmsFigWidth]{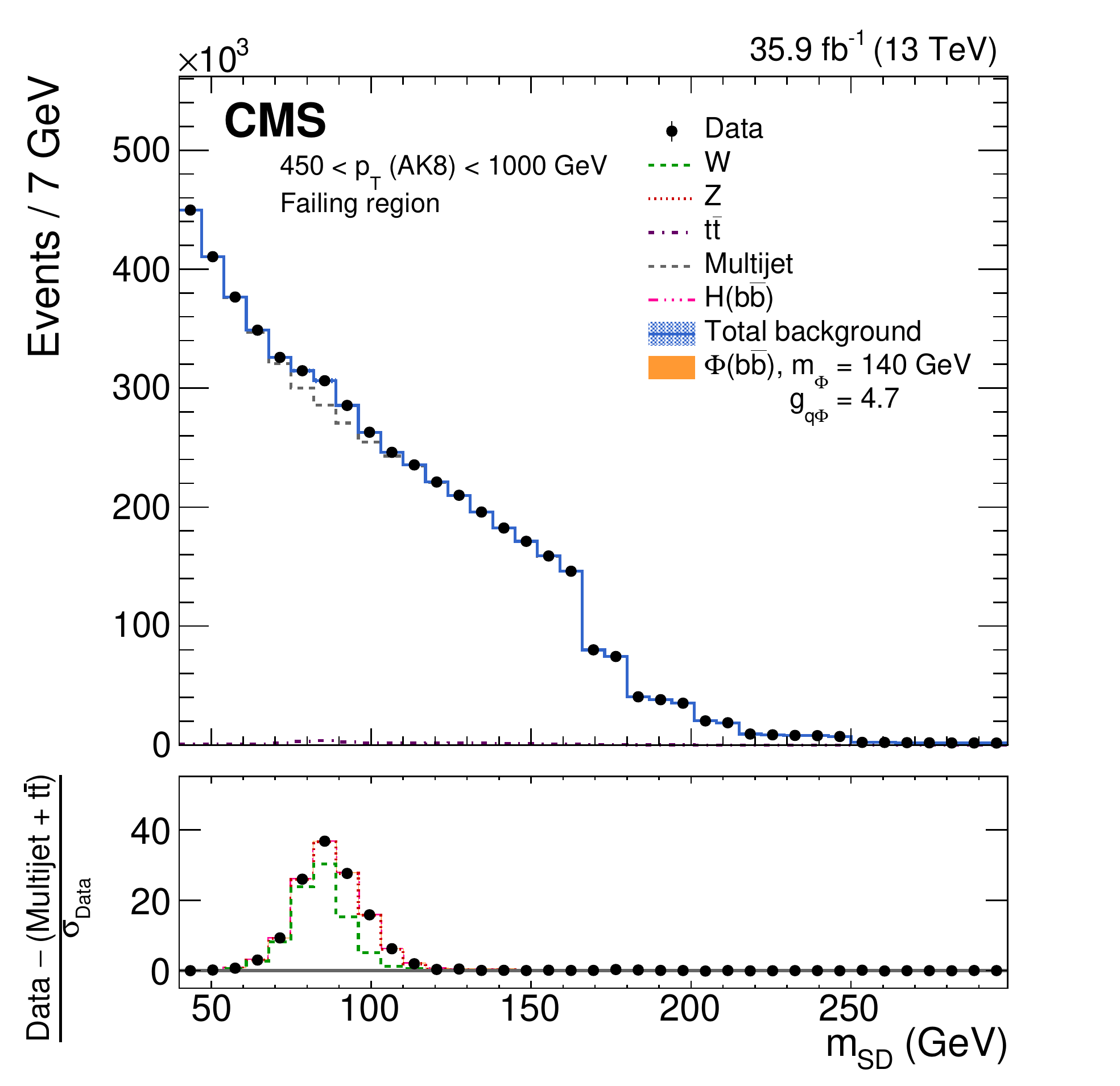}
    \includegraphics[width=\cmsFigWidth]{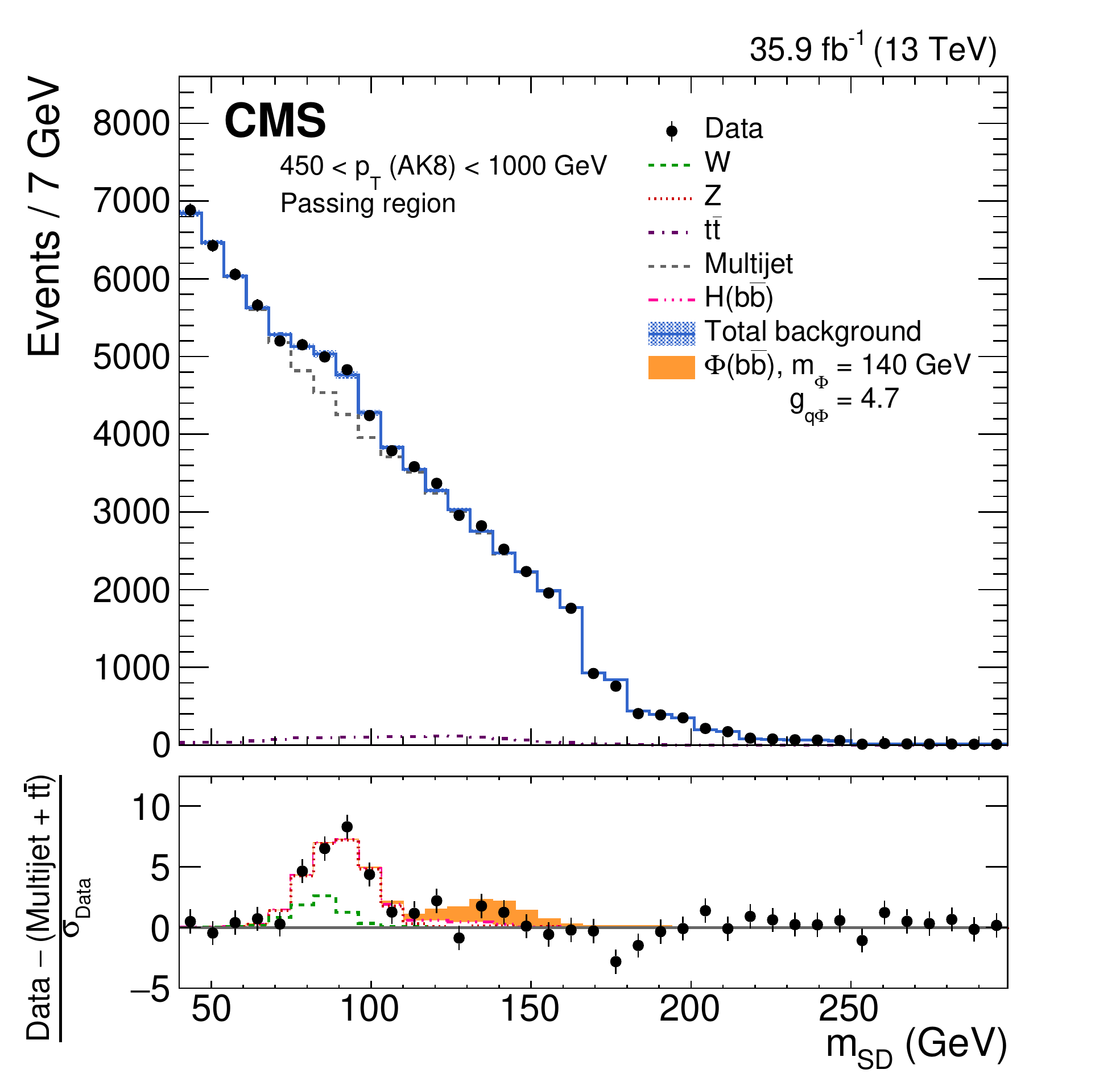}
    \caption{The observed and fitted background $\mSD$ distributions for the AK8 selection for the failing (left) and passing (right) regions, combining all the $\pt$ categories. The background fit is performed under the background-only hypothesis.
	A hypothetical $\Phi(\bbbar)$ signal at a mass of 140\GeV is also indicated.
      The features at 166, 180, 201, 215, and 250\GeV in the $\mSD$ distribution are due to the $\rho$ boundaries, which define different $\mSD$ ranges for each $\pt$ category. The shaded blue band shows the systematic uncertainty in the total background prediction. The bottom panel shows the difference between the data and the nonresonant background prediction, divided by the statistical uncertainty in the data.}
 \label{fig:resultsAK8}
 \end{figure*}

\begin{figure*}[hbtp]
\centering
    \includegraphics[width=\cmsFigWidth]{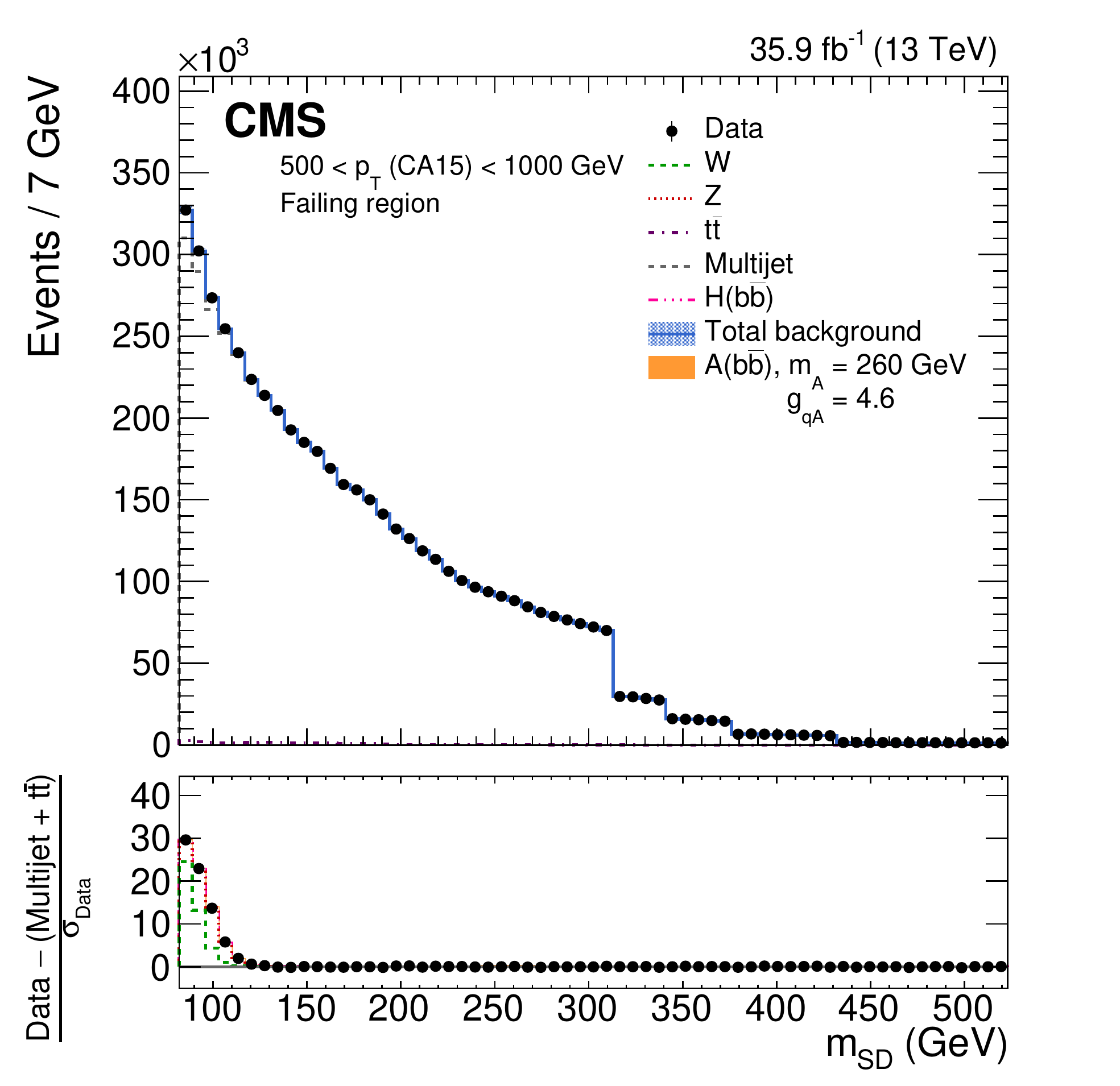}
    \includegraphics[width=\cmsFigWidth]{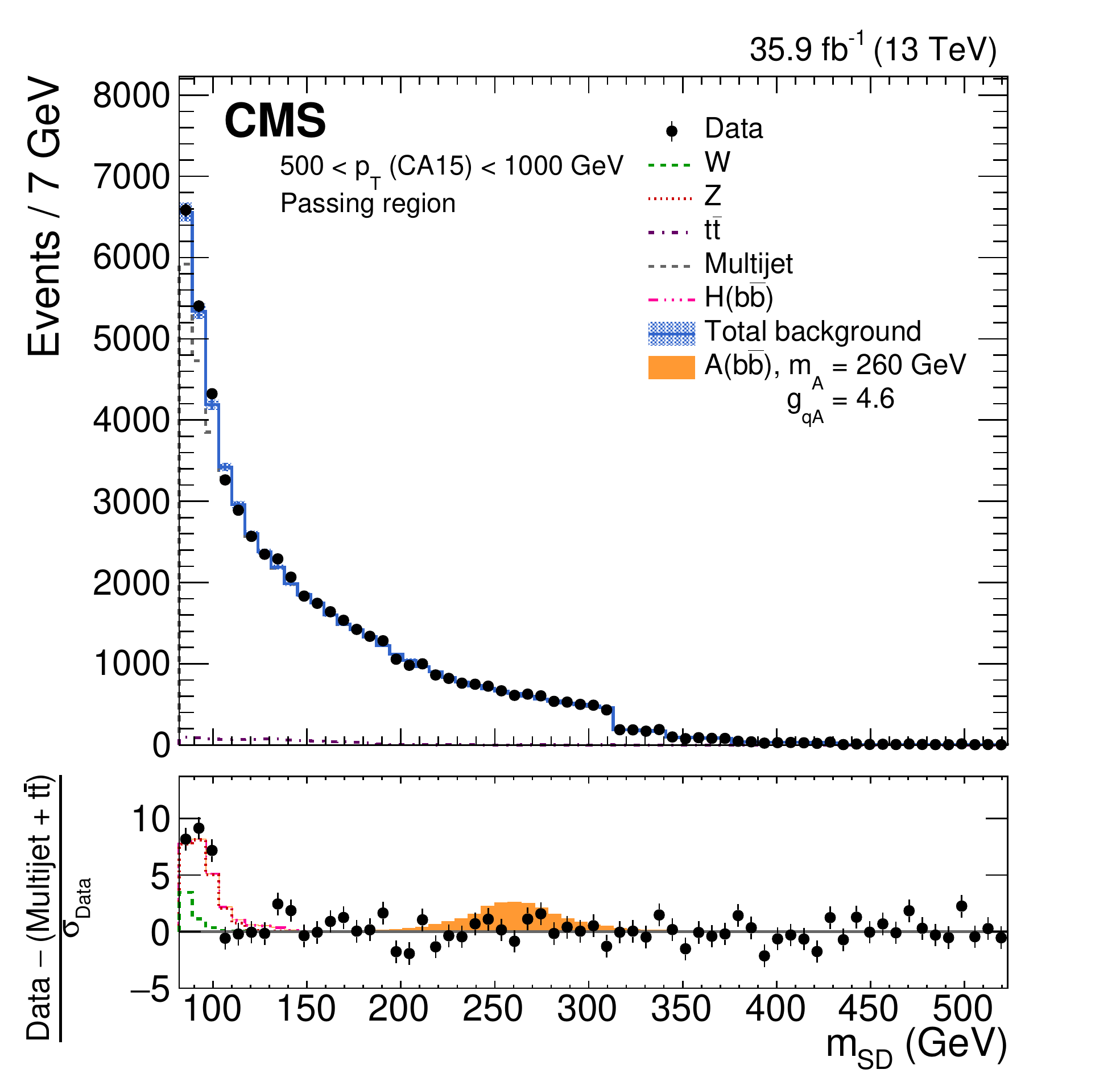}
 \caption{The observed and fitted background $\mSD$ distributions for the CA15 selection for the failing (\cmsLeft) and passing (\cmsRight) regions, combining all the $\pt$ categories. The background fit is performed under the background-only hypothesis. A hypothetical $\PA(\bbbar)$ signal at a mass of 260\GeV is also indicated. The features at 285, 313, 341, 376, and 432\GeV in the $\mSD$ distribution are due to the $\rho$ boundaries, which define different $\mSD$ ranges for each $\pt$ category. The shaded blue band shows the systematic uncertainty in the total background prediction. The bottom panel shows the difference between the data and the nonresonant background prediction, divided by the statistical uncertainty in the data.}
 \label{fig:resultsCA15}
 \end{figure*}

\begin{figure*}[hbtp]
  \centering
  \includegraphics[width=\cmsSmallFigWidth]{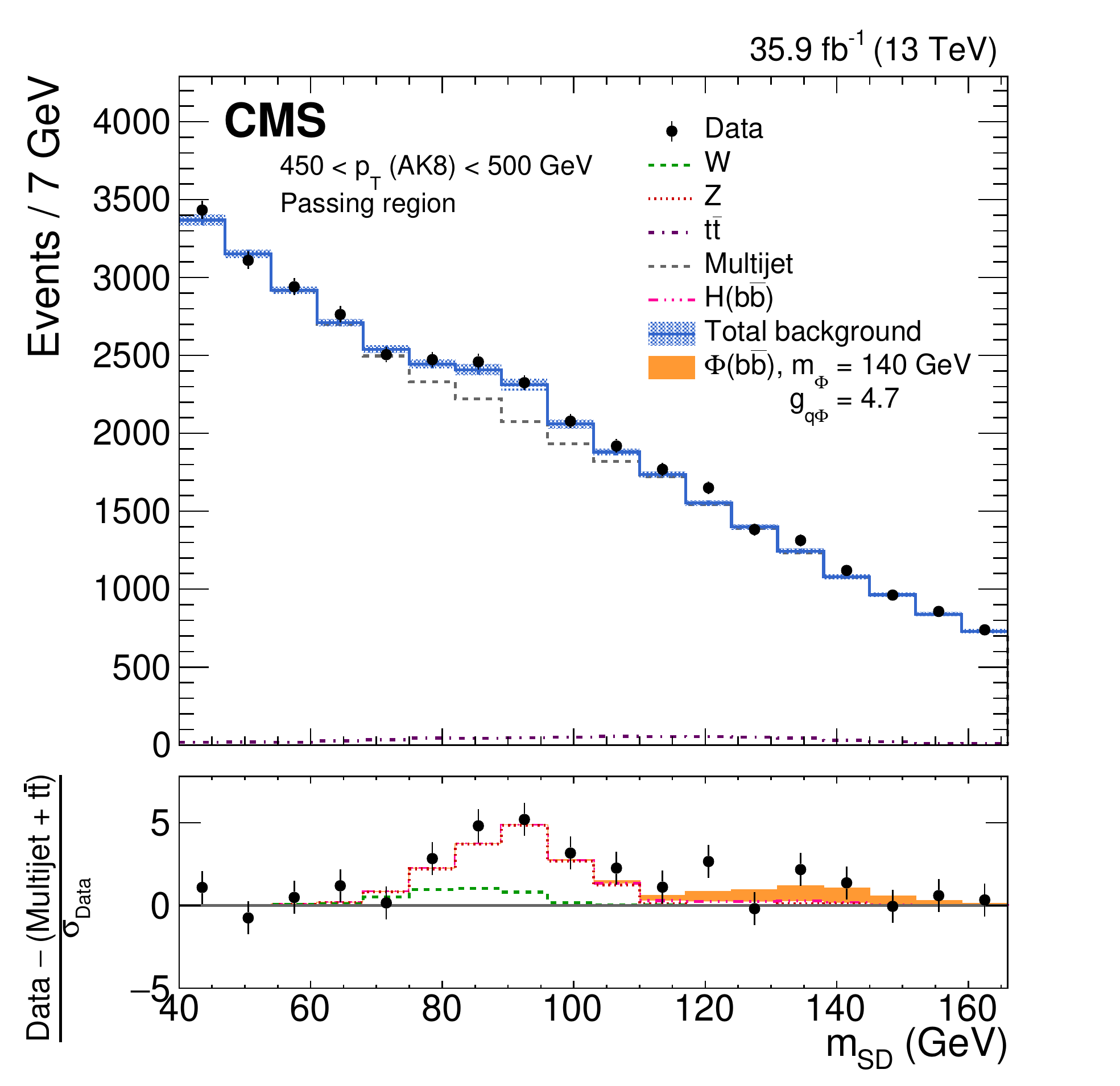}
  \includegraphics[width=\cmsSmallFigWidth]{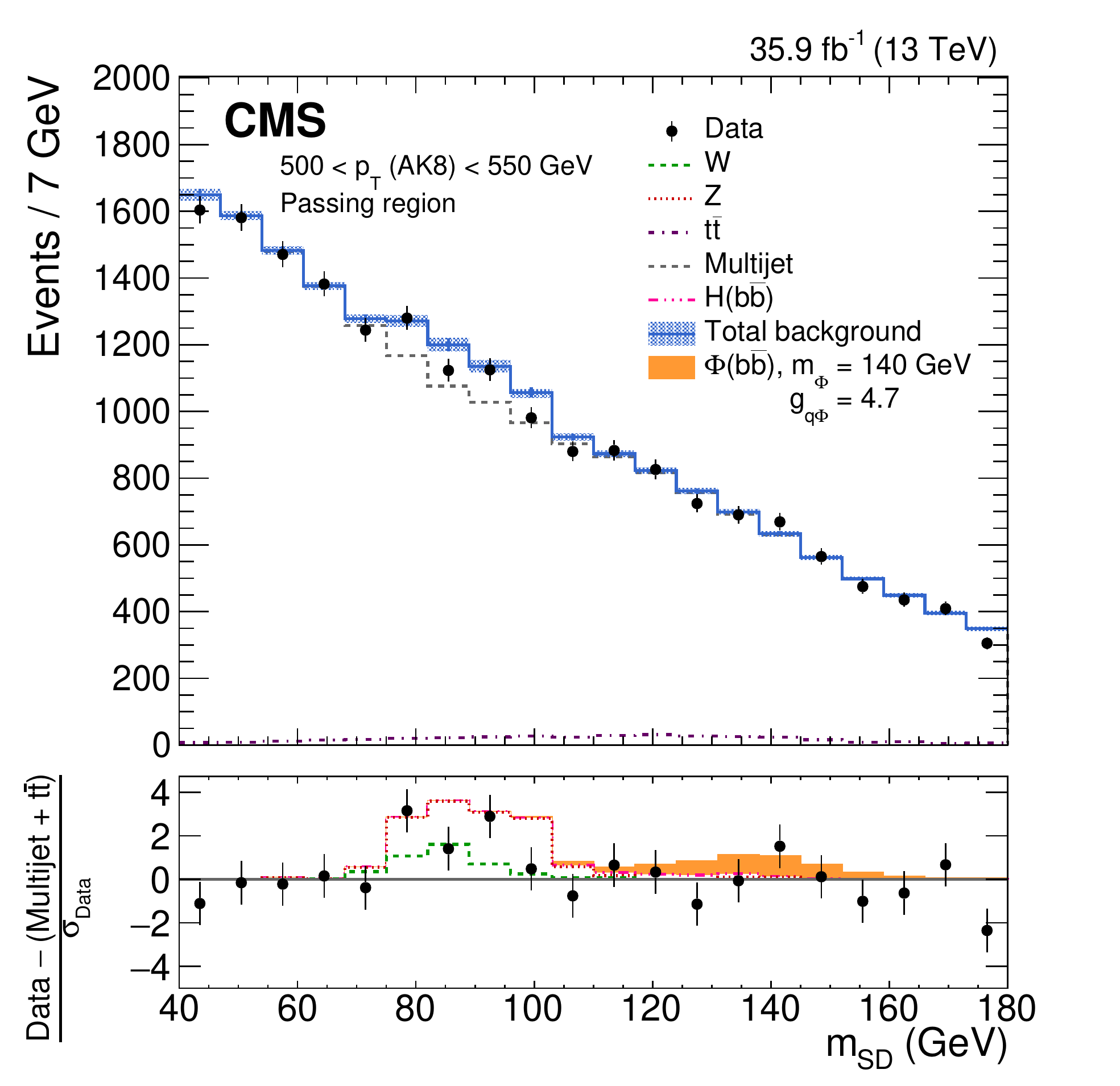}\\
  \includegraphics[width=\cmsSmallFigWidth]{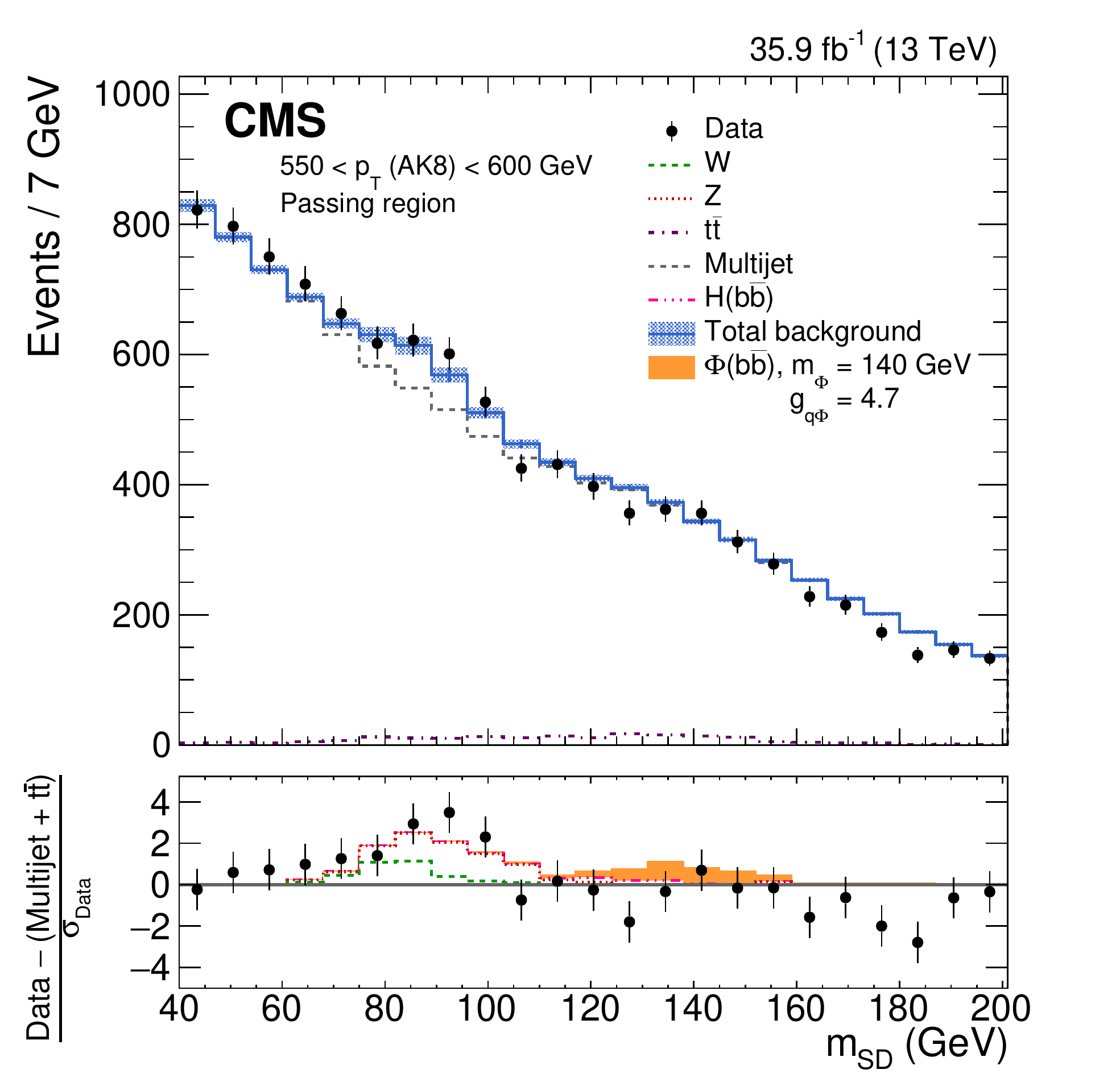}
  \includegraphics[width=\cmsSmallFigWidth]{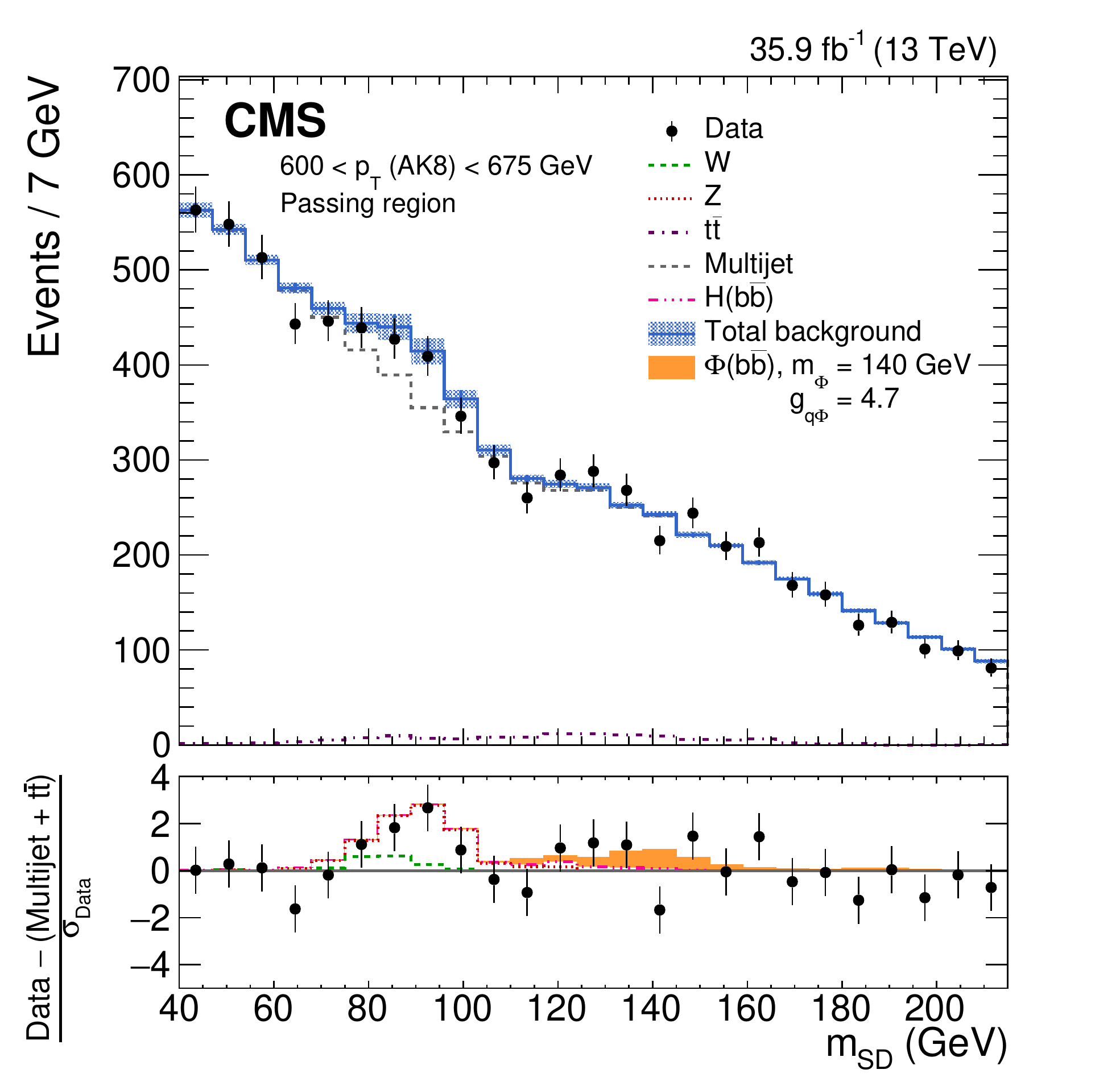}\\
  \includegraphics[width=\cmsSmallFigWidth]{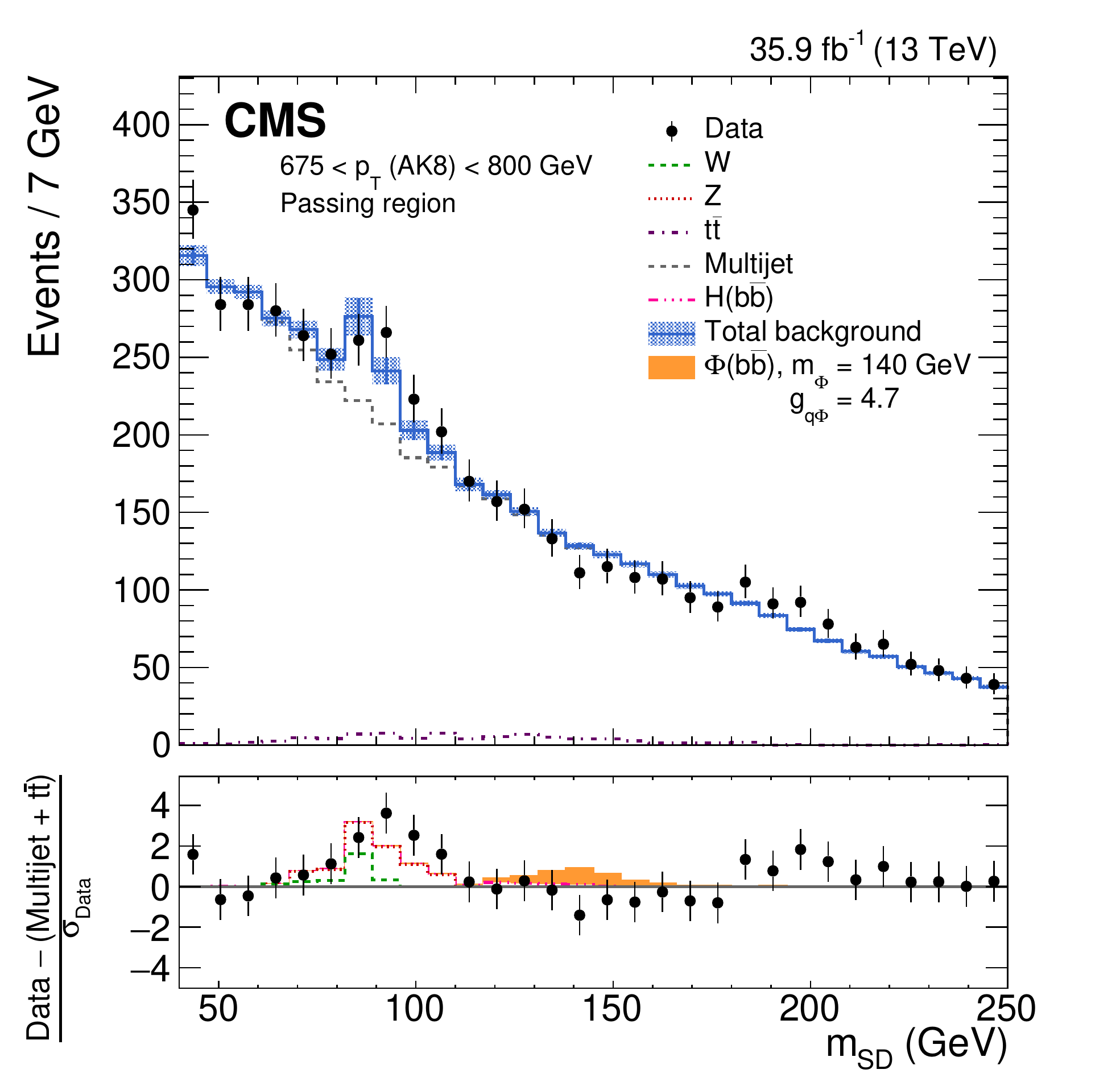}
  \includegraphics[width=\cmsSmallFigWidth]{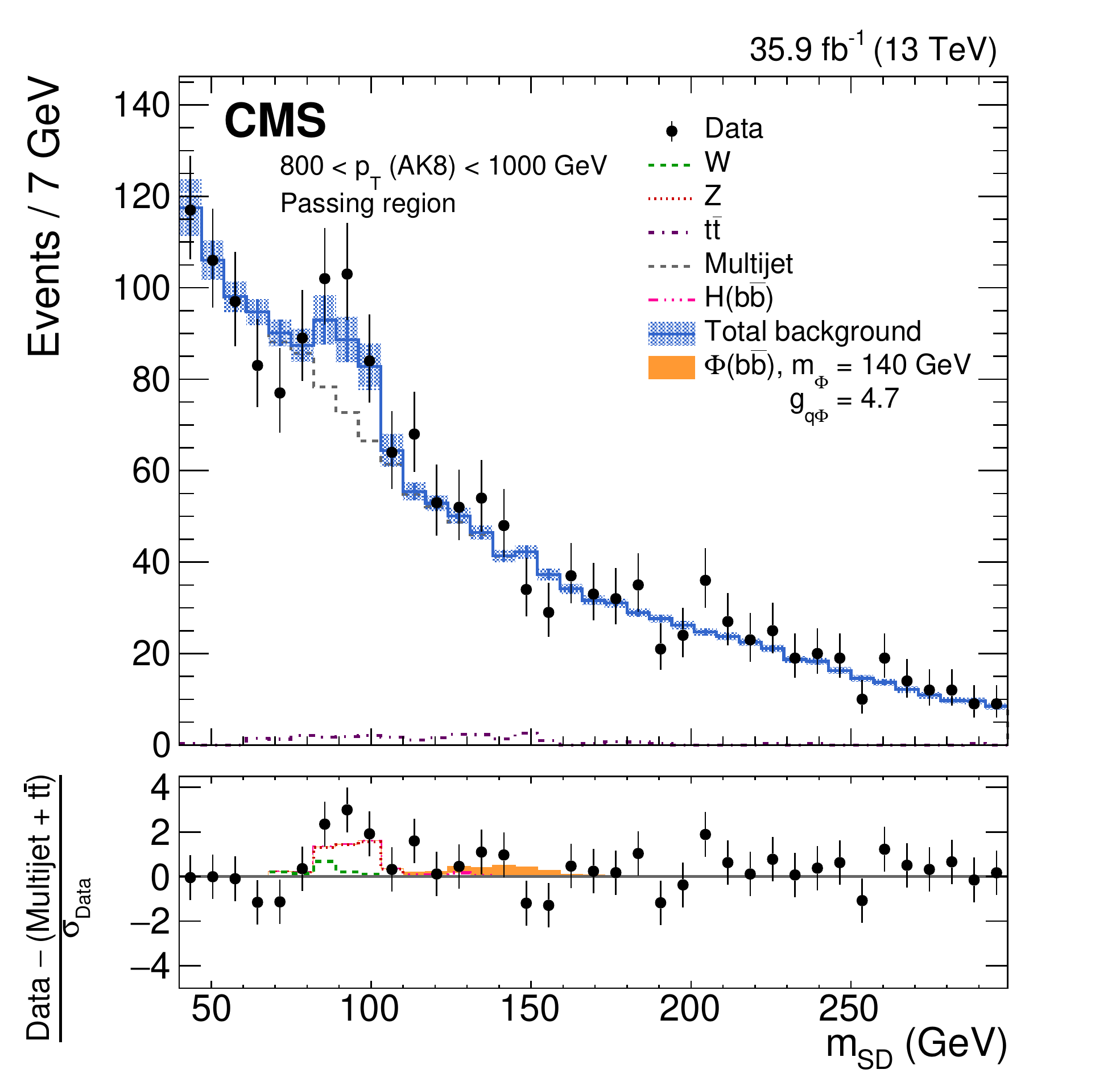}\\
  \caption{The observed and fitted background $\mSD$ distributions in each
    $\pt$ category for the AK8 selection in the passing regions. The fit is performed under the background-only hypothesis. A hypothetical $\Phi(\bbbar)$ signal at a mass of 140\GeV is also indicated. The shaded blue band shows the systematic uncertainty in the total background prediction. The bottom panel shows the difference between the data and the nonresonant background prediction, divided by the statistical uncertainty in the data.}
  \label{fig:resultsAK8_pt}
\end{figure*}

\begin{figure*}[hbtp]
  \centering
  \includegraphics[width=\cmsSmallFigWidth]{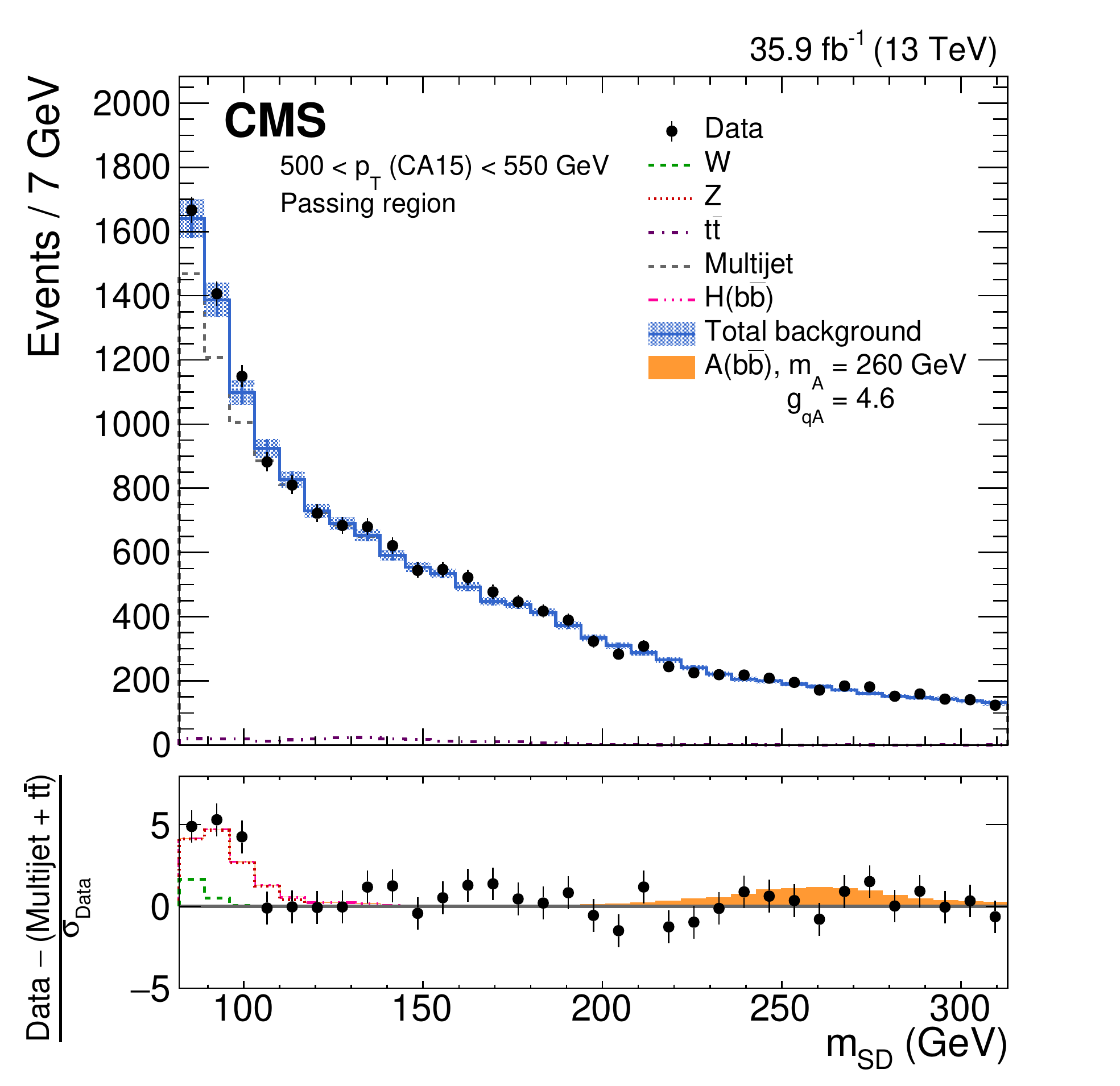}
  \includegraphics[width=\cmsSmallFigWidth]{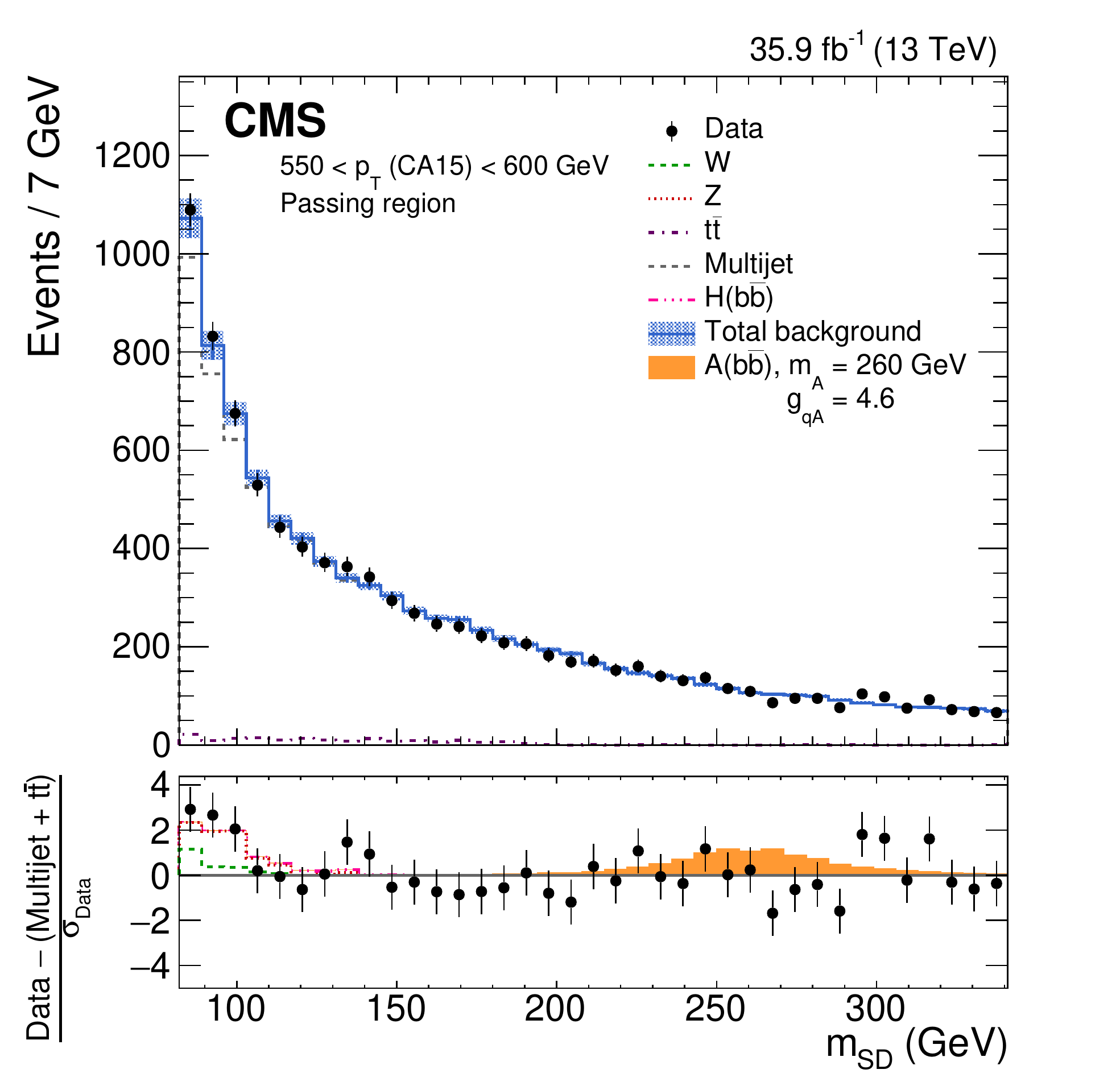}\\
  \includegraphics[width=\cmsSmallFigWidth]{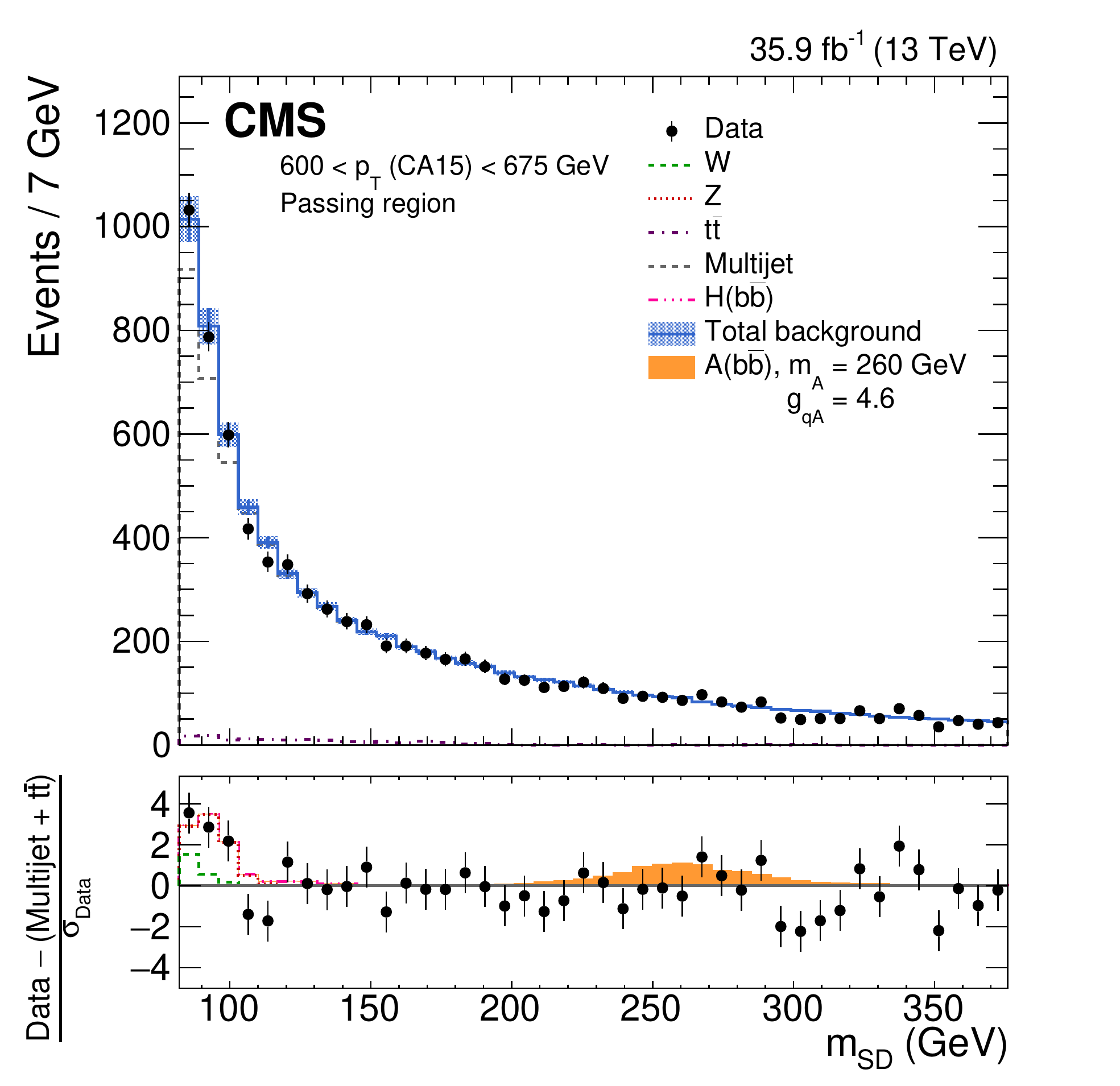}
  \includegraphics[width=\cmsSmallFigWidth]{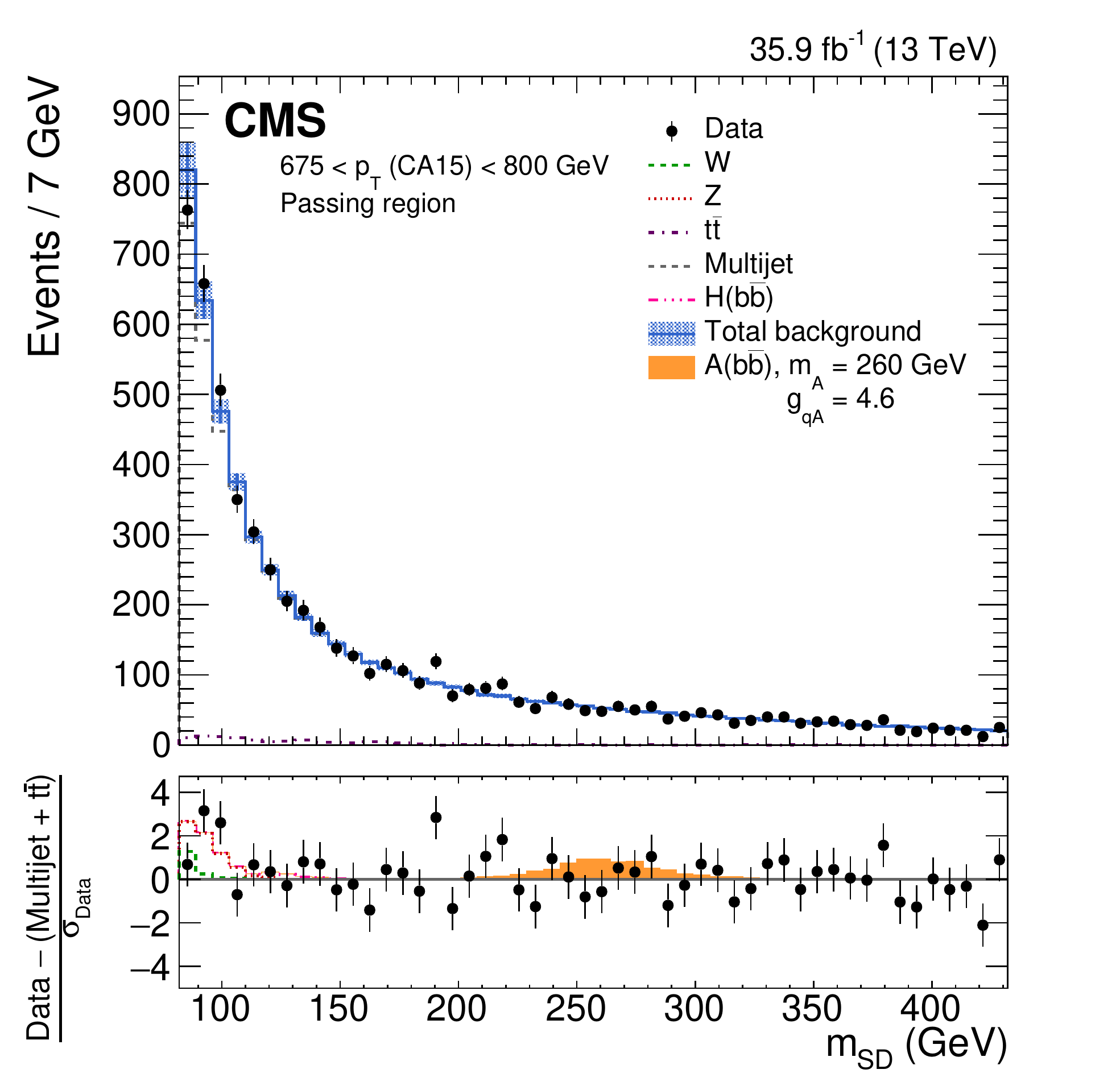}\\
  \includegraphics[width=\cmsSmallFigWidth]{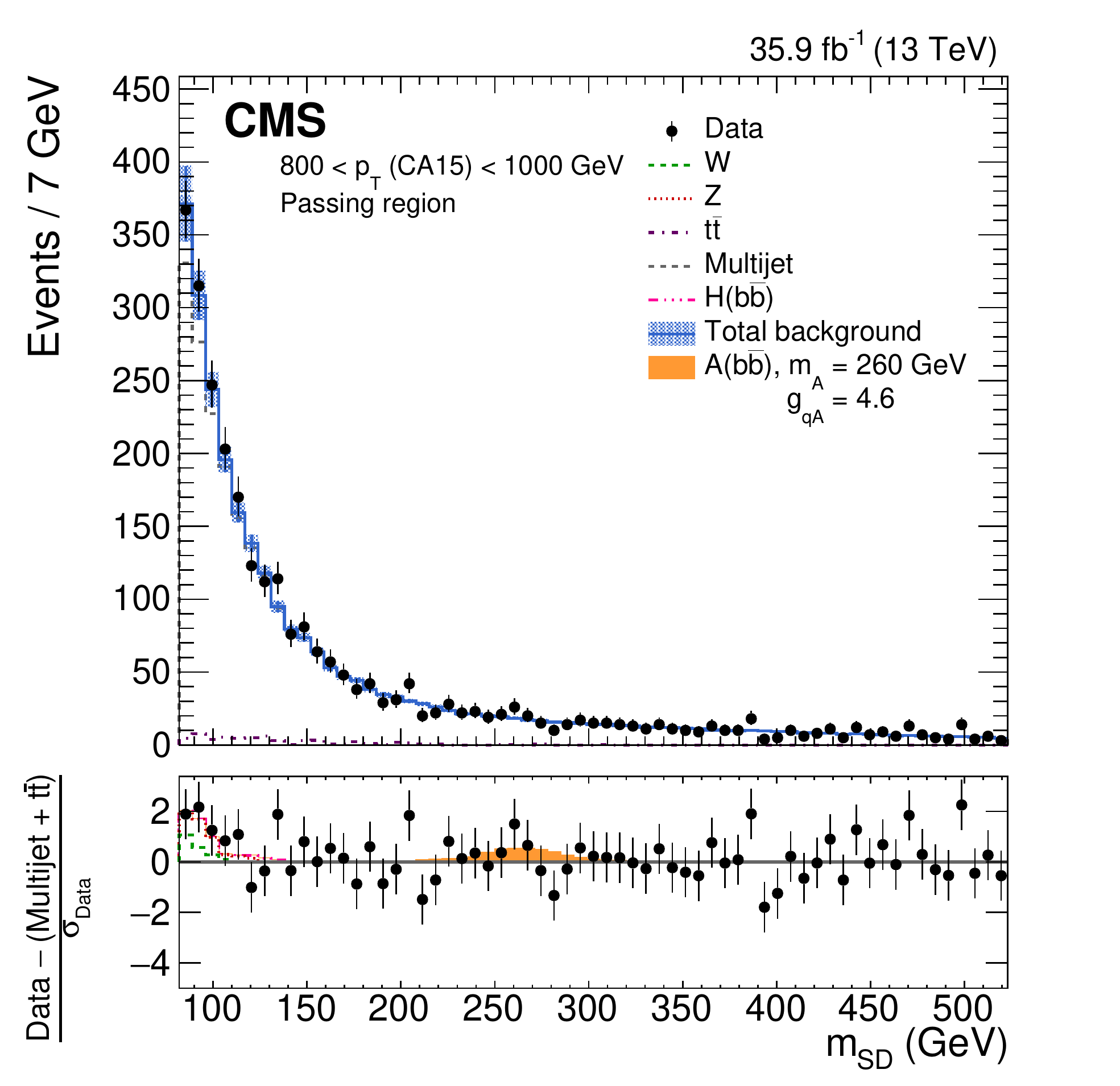}
  \caption{The observed and fitted background $\mSD$ distributions in each
    $\pt$ category for the CA15 selection in the passing regions. The fit is performed under the background-only hypothesis. A hypothetical $\PA(\bbbar)$ signal at a mass of 260\GeV is also indicated. The shaded blue band shows the systematic uncertainty in the total background prediction. The bottom panel shows the difference between the data and the nonresonant background prediction, divided by the statistical uncertainty in the data.}
  \label{fig:resultsCA15_pt}
\end{figure*}

Figures~\ref{fig:resultsAK8} and \ref{fig:resultsCA15} show the $\mSD$ distributions in the full data set for the passing and failing regions with fitted SM background
for AK8 and CA15 selections, respectively.
Note that the different $\rho$ boundaries define different $\mSD$ ranges for the AK8 and CA15 selections as well as within each $\pt$ category, giving rise to the features at 166, 180, 201, 215, and 250\GeV in Fig.~\ref{fig:resultsAK8} and at 285, 313, 341, 376, and 432\GeV in Fig.~\ref{fig:resultsCA15}.
The bottom panels of Figs.~\ref{fig:resultsAK8}--\ref{fig:resultsCA15_pt} show the difference between the data and the prediction from the nonresonant background, composed of the QCD multijet and $\ttbar$ processes, divided by the statistical uncertainty in the data. These highlight the agreement between the data and the contributions from $\PW$ and $\PZ$ boson production, which are clearly visible in the failing and passing regions, respectively.
The remaining $\PW$ boson contribution in the passing region is due to the misidentification of $\PW(\qqbar)$ decays.
No significant deviations from the background-only expectations are observed.
In Figs.~\ref{fig:resultsAK8_pt} and \ref{fig:resultsCA15_pt}, the $\mSD$ distributions are reported for each $\pt$ category for AK8 and CA15 jets, respectively.

In order to validate the background estimation method and associated systematic uncertainties, bias studies are performed on simulated samples and on the background-only fits. Pseudo-experiment data sets are generated, with and without the injection of signal events, and then fit with the signal plus background model.
No significant bias in the fitted signal strength is observed; specifically, the means of the differences between the fitted and injected signal strengths divided by the fitted uncertainty are found to be less than 15\%.

In addition, to validate the corrections and uncertainties related to the $\PW(\qqbar)$ and $\PZ(\qqbar)$ resonances, we perform a consistency check by directly measuring a combined signal strength for those contributions assuming the SM background-only hypothesis. We find agreement with the SM expectation within the measured uncertainties.

\section{Systematic uncertainties}
\label{sec:systematics}
The systematic uncertainties associated with the jet mass scale, the jet mass resolution, and the $\nddt$ selection efficiency are correlated among the $\PW$, $\PZ$, $\PH(\bbbar)$, and $\Phi(\bbbar)$ or $\PA(\bbbar)$ processes.
These uncertainties are estimated using an independent sample of merged $\PW$ jets in semileptonic $\ttbar$ events in data.

To select a sample of merged $\PW$ jets from semileptonic $\ttbar$ production, events are required to have an energetic muon with $\pt > 100\GeV$, $\ptmiss > 80\GeV$, a high-$\pt$ AK8  or CA15 jet with $\pt > 200\GeV$, and an additional jet separated from the AK8\,(CA15) jet by $\Delta R > 0.8\,(1.5)$.
Using the same $\nddt$ requirements that define the signal regions, we define samples with events that pass and fail the selection for merged $\PW$~boson jets in data and simulation.
A simultaneous fit to the two samples is performed in order to extract the selection efficiency of a merged $\PW$ jet in simulation and in data.
This is performed separately for AK8 and CA15 selections.
We measure the data-to-simulation scale factor for the $\nddt$ selection to be $0.99 \pm 0.04$ for AK8 jets and $0.97 \pm 0.06$ for CA15 jets.
The jet mass scales in data and simulation are found to be consistent within 1\%.
The jet mass resolution data-to-simulation scale factor is $1.08 \pm 0.09$ for AK8 jets and $0.99 \pm 0.08$ for CA15 jets.
As the semileptonic $\ttbar$ sample does not contain a large population of jets with very high $\pt$, an additional systematic uncertainty is included to account for the extrapolation to very high $\pt$ jets.
The jet mass scale uncertainty is allowed to vary in the signal extraction differently depending on the jet $\pt$, and ranges from 2\% at 450\GeV to 4\% at 1\TeV.

The efficiency of the double-$\cPqb$ tagger is measured in data and simulation in a sample enriched in $\bbbar$ pairs from gluon splitting~\cite{Sirunyan:2017ezt}.
Scale factors relating data and simulation are then computed and applied to the simulation.
The measured double-$\cPqb$ tagger efficiency scale factor is found to be $0.86 \pm 0.07$ for CA15 jets and $0.91 \pm 0.04$ for AK8 jets, where the uncertainty accounts for various systematic effects including the calibration of the jet probability tagger algorithm used in the method, the modeling of the track reconstruction efficiency, the modeling of $\cPqb$ quark fragmentation, and others~\cite{Sirunyan:2017ezt}.

The scale factors described above determine the initial distributions of the jet mass for the $\PW(\qqbar)$, $\PZ(\qqbar)$, $\PH(\bbbar)$, and $\Phi(\bbbar)$ or $\PA(\bbbar)$ processes and are further constrained in the fit to data by the presence of the $\PW$ and $\PZ$ resonances in the jet mass distribution.

To account for potential $\pt$-dependent deviations due to missing higher-order corrections, uncertainties are applied to the $\PW(\qqbar)$ and $\PZ(\qqbar)$ yields that are $\pt$ dependent and correlated per $\pt$ bin~\cite{Sirunyan:2017jix,Denner:2009gj,Denner:2011vu,Denner:2012ts,Kuhn:2005gv,Kallweit:2014xda,Kallweit:2015dum}.
An additional systematic uncertainty is included to account for potential differences between the $\PW$ and $\PZ$ higher-order corrections (EW $\PW/\PZ$ decorrelation)~\cite{Sirunyan:2017jix}.

Finally, additional systematic uncertainties are applied to the $\PW(\qqbar)$, $\PZ(\qqbar)$, $\ttbar$, $\PH(\bbbar)$, and $\Phi(\bbbar)$ or $\PA(\bbbar)$ yields to account for the uncertainties due to the jet energy scale and resolution~\cite{jec}, variations in the amount of pileup, the integrated luminosity determination~\cite{lumi}, and the limited simulation sample sizes. A quantitative summary of the systematic effects considered is shown in Table~\ref{tab:systematics}.

\begin{table*}
\centering
\topcaption{
Summary of the systematic uncertainties affecting the signal, $\PW$ and $\PZ+$~jets processes.  Instances where the uncertainty does not apply are indicated by a dash. The reported percentages reflect a one standard deviation effect on the product of acceptance and efficiency of each process. For the uncertainties related to the jet mass scale and resolution, which affect the mass distribution shapes, the reported percentages reflect a one standard deviation effect on the nominal jet mass.}
\label{tab:systematics}
\resizebox{\cmsTabWidth}{!}{
\begin{scotch}{lcccc}
Uncertainty source & \multicolumn{4}{c}{Process}  \\
& $\PW$ or $\PZ$ (AK8) & $\PW$ or $\PZ$ (CA15) & $\Phi$ or $\PA$ (AK8) & $\Phi$ or $\PA$ (CA15)\vspace{\cmsTabSkip}\\
Integrated luminosity         & 2.5\% & 2.5\% & 2.5\% & 2.5\% \\
Trigger efficiency & 2\% & 2\% & 2\% & 2\% \\
Pileup    & $<$1\% & $<$1\%  & $<$1\% & $<$1\% \\
$\nddt$ selection efficiency & 4.3\% & 6\% & 4.3\% & 6\% \\
Double-$\cPqb$ tag  & 4\% (\PZ)  & 8\% (\PZ) & 4\%  & 8\% \\
Jet energy scale / resolution  & 5--15\% & 5--15\%  & 5--15\%  & 5--15\%  \\
Jet mass resolution & 8\% & 8\%  & 8\% & 8\%  \\
Jet mass scale $\left(\%\,/\,(\pt\,\mathrm{[GeV]}\,/\,100)\right)$  & 0.4\% & 1\%  & 0.4\%  & 1\% \\
Simulation sample size        & 2--25\% & 2--25\% & 4--20\% & 4--20\% \\
NLO QCD corrections &  10\% & 10\% & \NA & \NA\\
NLO EW corrections &  15--35\% &  15--35\% & \NA & \NA\\
NLO EW $\PW/\PZ$ decorrelation & 5--15\% & 5--15\% & \NA & \NA \\
\end{scotch}}
\end{table*}

\section{Results}
\label{sec:results}

The search results are interpreted in the context of the scalar and pseudoscalar signal models described in Section~\ref{sec:intro}. The signals are modeled using MC simulation.
For the search with AK8 (CA15) jets, a binned maximum likelihood fit to the observed $\mSD$ distributions in the range 40 to 201 (82 to 399)\GeV with a 7\GeV bin width is performed using the sum of the signal, $\PH(\bbbar)$, $\PW$, $\PZ$, $\ttbar$, and QCD multijet contributions.
The fit is performed simultaneously in the passing and failing regions of the six (five) $\pt$ categories within $450 (500) <\pt <1000\GeV$ for AK8 (CA15) jets, as well as in the passing and failing components of the $\ttbar$-enriched control region.

The chosen test statistic, used to determine how signal- or background-like the data are, is based on the profile likelihood ratio~\cite{LHCCLs} using the \CLs criterion~\cite{CLS1,CLS2}.
Systematic uncertainties are incorporated into the analysis via nuisance parameters and treated according to the frequentist paradigm.
Upper limits at 95\% confidence level (\CL) are obtained using asymptotic formulae~\cite{LHCCLs,Khachatryan:2014jba,Cowan:2010js}.

\begin{figure}[hbtp]
\centering
\includegraphics[width=\cmsFigWidth]{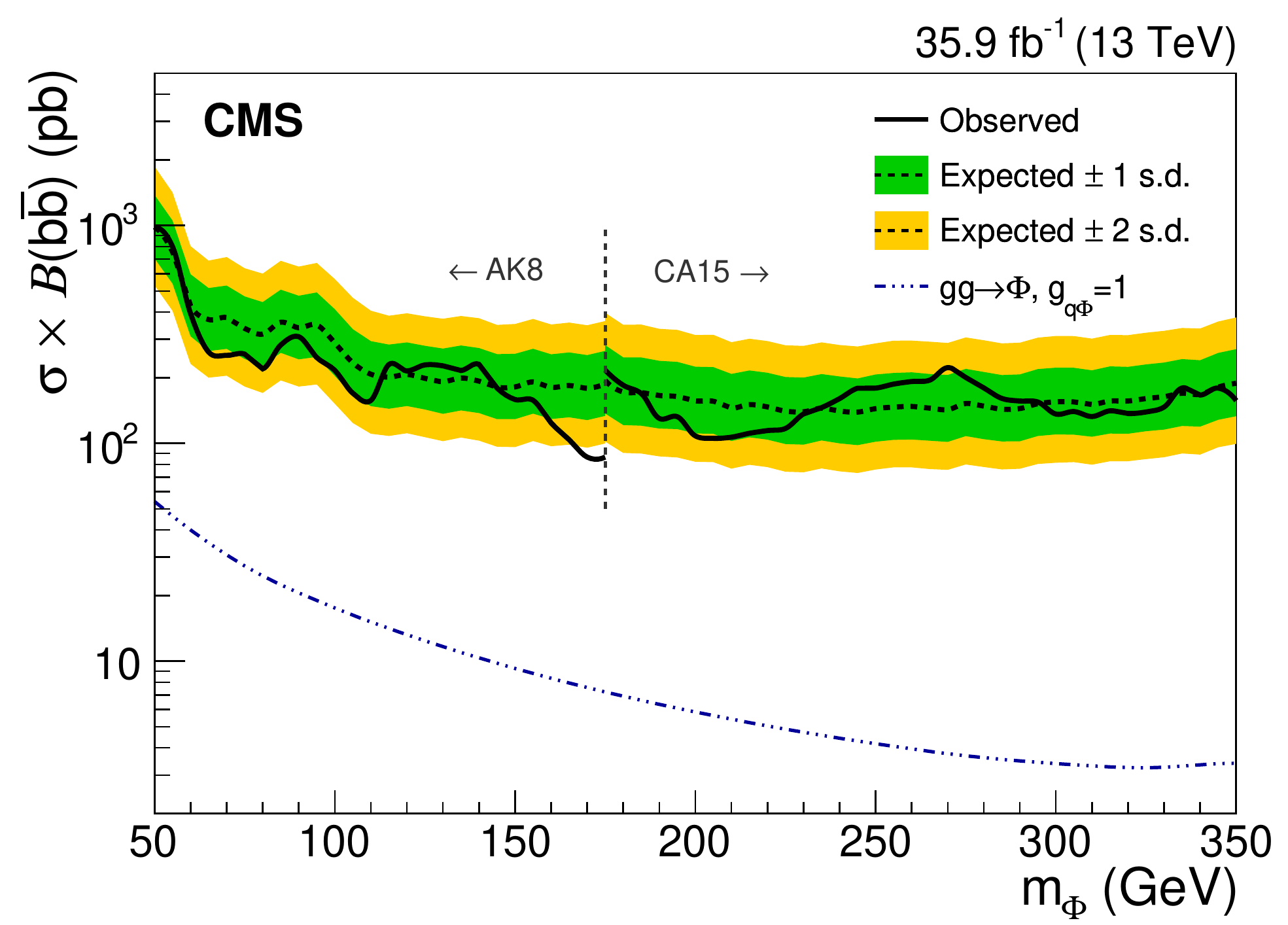}
\includegraphics[width=\cmsFigWidth]{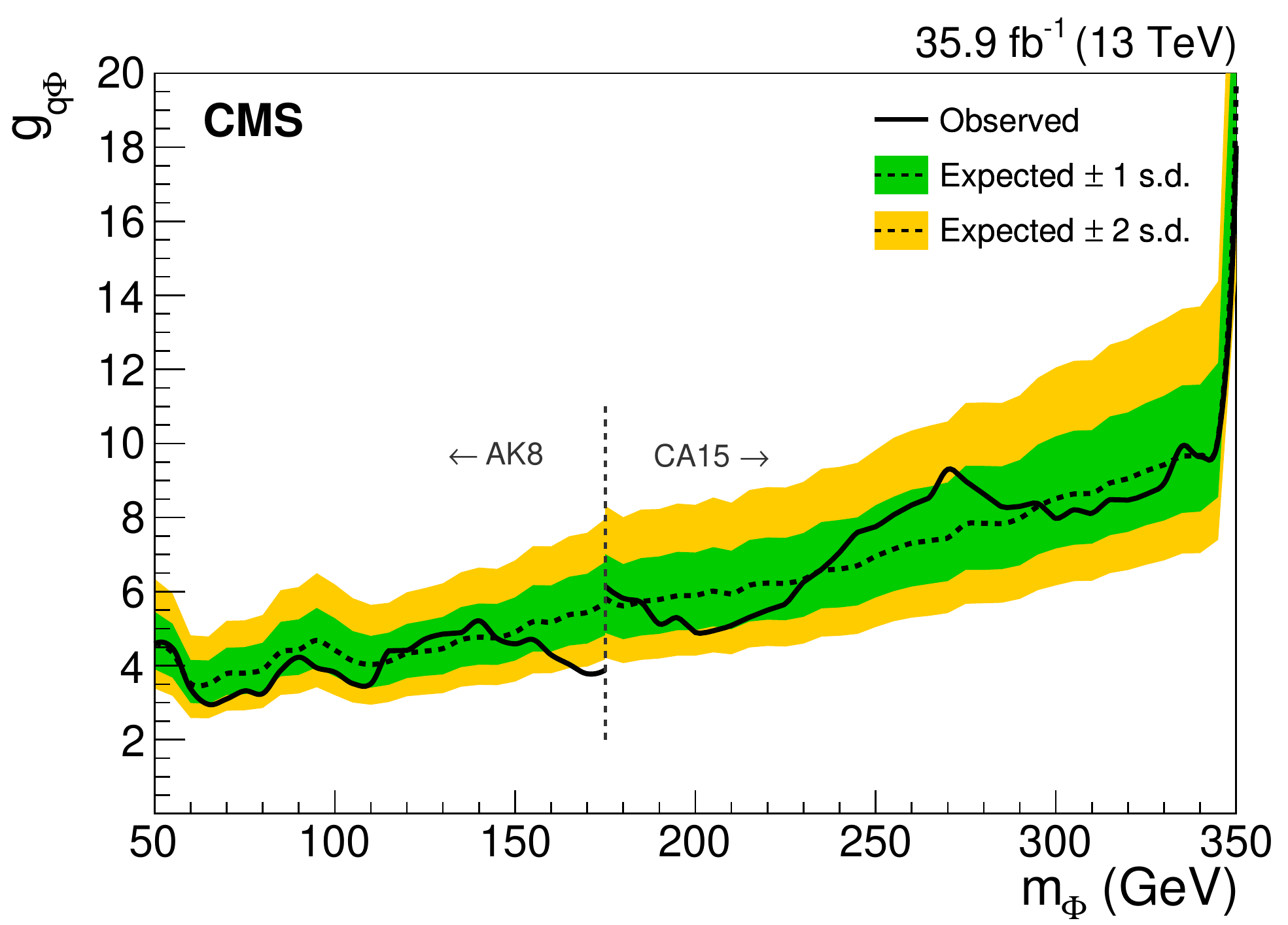}
        \caption{Upper limits at 95\% \CL on the product of the $\Phi$ production
          cross section and the branching fraction to $\bbbar$ (\cmsLeft) and on $\gqphi$ (\cmsRight), as a function of the resonance mass $m_{\Phi}$. The blue dash-dotted line indicates the theoretical scalar production cross section assuming $\gqphi=1$ as a chosen benchmark~\cite{Boveia:2016mrp}. The vertical line at 175\GeV corresponds to the transition between the AK8 and CA15 jet selections. }
        \label{fig:LimitsPhi}
 \end{figure}

\begin{figure}[hbtp]
\centering
\includegraphics[width=\cmsFigWidth]{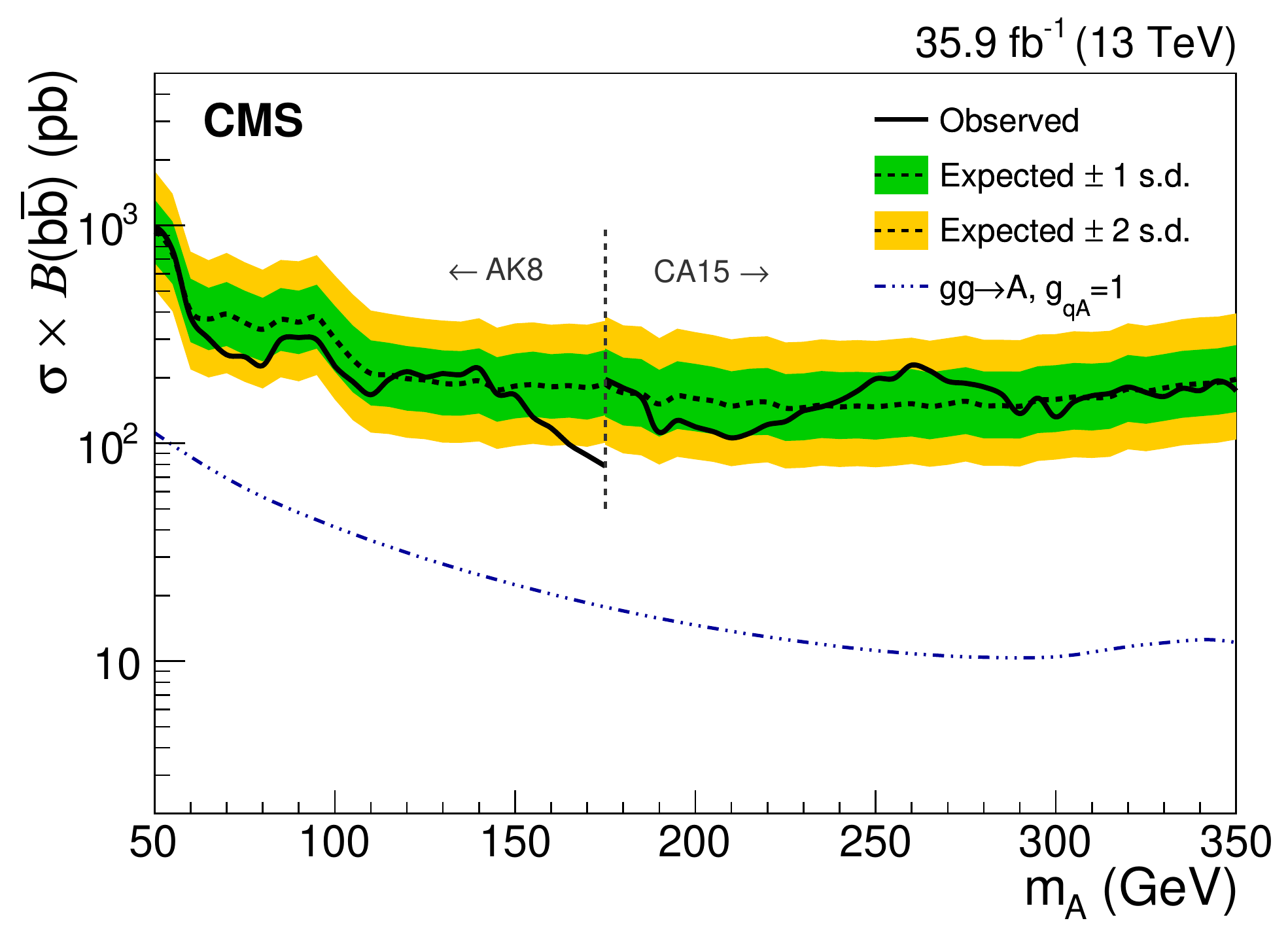}
\includegraphics[width=\cmsFigWidth]{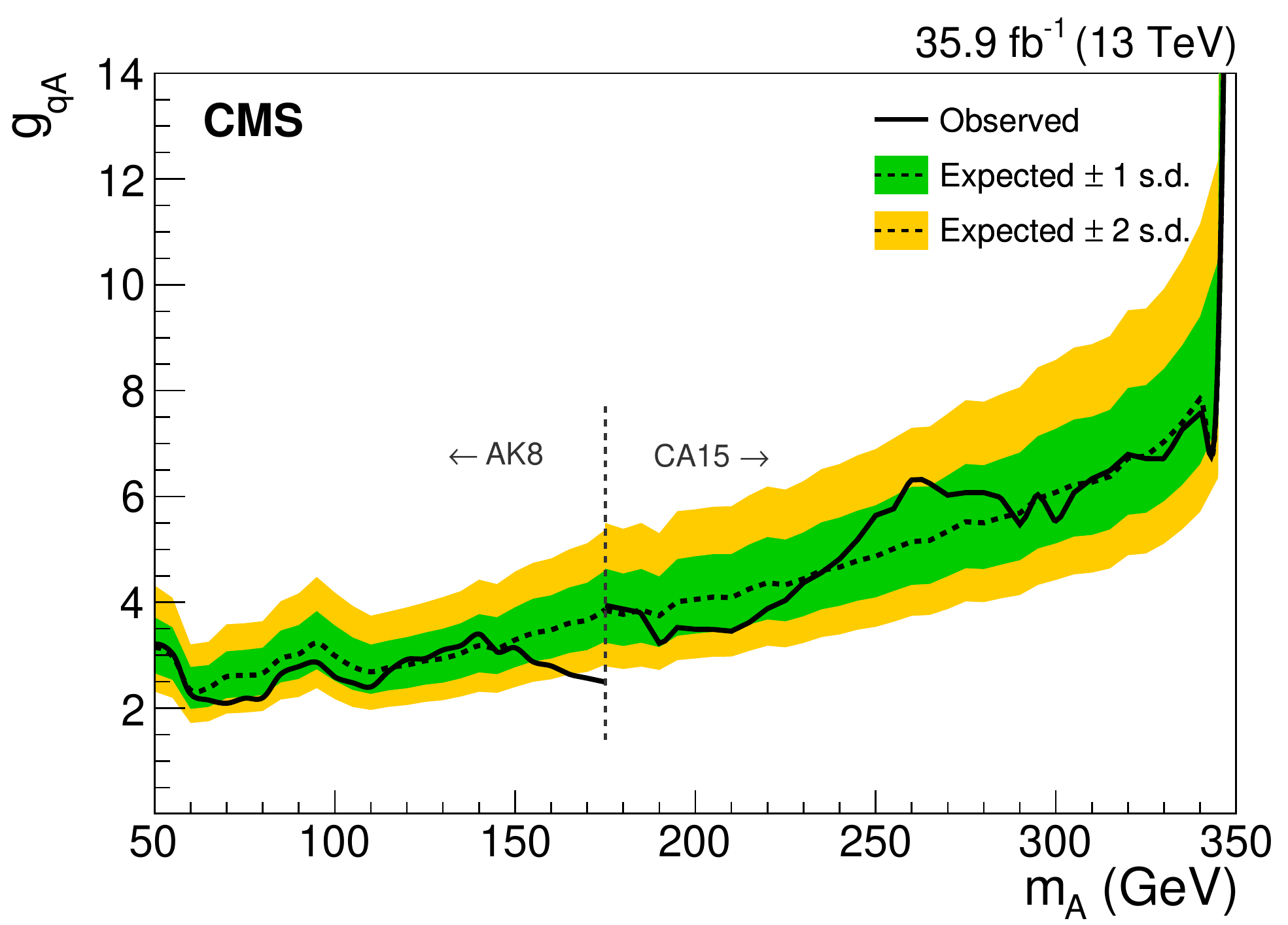}
        \caption{Upper limits at 95\% \CL on the product of the $\PA$ production
          cross section and the branching fraction to $\bbbar$ (\cmsLeft) and on $\gqa$ (\cmsRight), as a function of the resonance mass $m_{\PA}$. The blue dash-dotted line indicates the theoretical pseudoscalar production cross section assuming $\gqa=1$ as a chosen benchmark~\cite{Boveia:2016mrp}. The vertical line at 175\GeV corresponds to the transition between the AK8 and CA15 jet selections. }
        \label{fig:LimitsA}
 \end{figure}

The 95\% \CL upper limits on the $\Phi(\bbbar)$ and $\PA(\bbbar)$ production as a function
of resonance mass are shown in Figs.~\ref{fig:LimitsPhi} and~\ref{fig:LimitsA}, respectively.
Based on the expected sensitivity, the AK8 jet selection is used for signal masses below 175\GeV, and the CA15 jet selection is used above.
We exclude $\Phi$ or $\PA$ production with a product of the cross section and branching fraction ($\sigma \mathcal{B}(\bbbar)$) as low as 79 or $86\unit{pb}$, respectively, at a resonance mass of 175\GeV.
The exclusions are converted to upper limits on the coupling $\gqphi$ for the scalar model and the coupling $\gqa$ for the pseudoscalar model.
The abrupt loss in sensitivity to the coupling constants for resonance masses greater than $2m_\cPqt$ is because the branching fraction to $\bbbar$ falls steeply as the decay to $\ttbar$ becomes kinematically allowed.
For a resonance mass of 175\GeV, the exclusion corresponds to an upper limit on $\gqphi$ or $\gqa$ of 3.9 or 2.5, respectively.

For the search with AK8 jets, the maximum local significance~\cite{pvalue} corresponds to 0.5 standard deviations from the background-only expectation at a $\Phi(\bbbar)$ mass of 140\GeV. The hypothetical $\Phi(\bbbar)$ signal is indicated in Figs.~\ref{fig:resultsAK8} and \ref{fig:resultsAK8_pt} with $\gqphi=4.7$, which is equivalent to the 95\% \CL upper limit. Similarly, for the CA15 search, the maximum local significance is 1.2 standard deviations at an $\PA(\bbbar)$ mass of 260\GeV. The hypothetical $\PA(\bbbar)$ signal is indicated in Figs.~\ref{fig:resultsCA15} and \ref{fig:resultsCA15_pt} with $\gqa=4.6$, which is equivalent to the 95\% \CL upper limit. The largest downward fluctuation in the limits occurs at an $\PA(\bbbar)$ mass of 175\GeV in the AK8 search, corresponding to a local significance of $-2.9$ standard deviations and a global significance~\cite{pvalue}, calculated over the probed mass range (50--350\GeV), of approximately $-1.7$ standard deviations. A corresponding deficit is not seen in CA15 search, as the events used in the AK8 and CA15 searches are largely independent; approximately 20 (37)\% of events in the CA15 search are selected in the AK8 search, while conversely, approximately 37\% of events in the AK8 search are selected in the CA15 search.

\section{Summary}
\label{sec:summary}

A search for a low-mass resonance decaying into a bottom quark-antiquark pair and reconstructed as a single wide jet
has been presented, using a data set of proton-proton collisions at \mbox{$\sqrt{s}=13\TeV$} corresponding to an integrated luminosity of 35.9\fbinv.
Dedicated substructure and double-$\cPqb$ tagging techniques were employed to identify jets containing a resonance candidate over a smoothly falling soft-drop jet mass distribution in data.
No significant excess above the standard model prediction was observed for signal masses between 50--350\GeV. Upper limits at 95\% confidence level are set on the product of the resonance production cross section and the branching fraction to bottom quark-antiquark pairs, as well as on the coupling $\gqphi$ ($\gqa$) of a scalar (pseudoscalar) boson decaying to quarks.
The search excludes the production through gluon fusion of a scalar (pseudoscalar) decaying to $\bbbar$ with a product of the cross section and branching fraction as low as 79 (86)\unit{pb} at a resonance mass of 175\GeV, corresponding to an upper limit on $\gqphi$ ($\gqa$) of 3.9 (2.5). This constitutes the first LHC constraint on exotic bottom quark-antiquark resonances below 325\GeV.

\begin{acknowledgments}
We congratulate our colleagues in the CERN accelerator departments for the excellent performance of the LHC and thank the technical and administrative staffs at CERN and at other CMS institutes for their contributions to the success of the CMS effort. In addition, we gratefully acknowledge the computing centers and personnel of the Worldwide LHC Computing Grid for delivering so effectively the computing infrastructure essential to our analyses. Finally, we acknowledge the enduring support for the construction and operation of the LHC and the CMS detector provided by the following funding agencies: BMBWF and FWF (Austria); FNRS and FWO (Belgium); CNPq, CAPES, FAPERJ, FAPERGS, and FAPESP (Brazil); MES (Bulgaria); CERN; CAS, MoST, and NSFC (China); COLCIENCIAS (Colombia); MSES and CSF (Croatia); RPF (Cyprus); SENESCYT (Ecuador); MoER, ERC IUT, and ERDF (Estonia); Academy of Finland, MEC, and HIP (Finland); CEA and CNRS/IN2P3 (France); BMBF, DFG, and HGF (Germany); GSRT (Greece); NKFIA (Hungary); DAE and DST (India); IPM (Iran); SFI (Ireland); INFN (Italy); MSIP and NRF (Republic of Korea); MES (Latvia); LAS (Lithuania); MOE and UM (Malaysia); BUAP, CINVESTAV, CONACYT, LNS, SEP, and UASLP-FAI (Mexico); MOS (Montenegro); MBIE (New Zealand); PAEC (Pakistan); MSHE and NSC (Poland); FCT (Portugal); JINR (Dubna); MON, RosAtom, RAS, RFBR, and NRC KI (Russia); MESTD (Serbia); SEIDI, CPAN, PCTI, and FEDER (Spain); MOSTR (Sri Lanka); Swiss Funding Agencies (Switzerland); MST (Taipei); ThEPCenter, IPST, STAR, and NSTDA (Thailand); TUBITAK and TAEK (Turkey); NASU and SFFR (Uxkraine); STFC (United Kingdom); DOE and NSF (USA).

\hyphenation{Rachada-pisek} Individuals have received support from the Marie-Curie program and the European Research Council and Horizon 2020 Grant, contract No. 675440 (European Union); the Leventis Foundation; the A. P. Sloan Foundation; the Alexander von Humboldt Foundation; the Belgian Federal Science Policy Office; the Fonds pour la Formation \`a la Recherche dans l'Industrie et dans l'Agriculture (FRIA-Belgium); the Agentschap voor Innovatie door Wetenschap en Technologie (IWT-Belgium); the F.R.S.-FNRS and FWO (Belgium) under the ``Excellence of Science - EOS" - be.h project n. 30820817; the Ministry of Education, Youth and Sports (MEYS) of the Czech Republic; the Lend\"ulet (``Momentum") Program and the J\'anos Bolyai Research Scholarship of the Hungarian Academy of Sciences, the New National Excellence Program \'UNKP, the NKFIA research grants 123842, 123959, 124845, 124850 and 125105 (Hungary); the Council of Science and Industrial Research, India; the HOMING PLUS program of the Foundation for Polish Science, cofinanced from European Union, Regional Development Fund, the Mobility Plus program of the Ministry of Science and Higher Education, the National Science Center (Poland), contracts Harmonia 2014/14/M/ST2/00428, Opus 2014/13/B/ST2/02543, 2014/15/B/ST2/03998, and 2015/19/B/ST2/02861, Sonata-bis 2012/07/E/ST2/01406; the National Priorities Research Program by Qatar National Research Fund; the Programa Estatal de Fomento de la Investigaci{\'o}n Cient{\'i}fica y T{\'e}cnica de Excelencia Mar\'{\i}a de Maeztu, grant MDM-2015-0509 and the Programa Severo Ochoa del Principado de Asturias; the Thalis and Aristeia programs cofinanced by EU-ESF and the Greek NSRF; the Rachadapisek Sompot Fund for Postdoctoral Fellowship, Chulalongkorn University and the Chulalongkorn Academic into Its 2nd Century Project Advancement Project (Thailand); the Welch Foundation, contract C-1845; and the Weston Havens Foundation (USA).

\end{acknowledgments}
\bibliography{auto_generated}

\providecommand{\href}[2]{#2}\begingroup\raggedright\begin{thebibliography}{10}%
\makeatletter
\providecommand{\hrefCMSnoop }[0]{\@secondoftwo}%
\makeatother
\providecommand{\doi}{\texttt{doi:}\begingroup \urlstyle{tt}\Url}

\bibitem{Abercrombie:2015wmb}
\hrefCMSnoop {}{Abercrombie {et~al.}, ``Dark matter benchmark models for early
  {LHC} {R}un-2 searches: Report of the {ATLAS}/{CMS} dark matter forum'',}
  (2015).
\href{http://www.arXiv.org/abs/1507.00966}{\texttt{arXiv:1507.00966}}.

\bibitem{Buckley:2014fba}
\hrefCMSnoop {}{M.~R. Buckley, D.~Feld, and D.~Goncalves, ``Scalar simplified
  models for dark matter'',} \textit{ Phys. Rev. D} \textbf{ 91} (2015) 015017,
  \href{http://dx.doi.org/10.1103/PhysRevD.91.015017}{\doi{10.1103/PhysRevD.91.015017}},
\href{http://www.arXiv.org/abs/1410.6497}{\texttt{arXiv:1410.6497}}.

\bibitem{Harris:2014hga}
\hrefCMSnoop {}{P.~Harris, V.~V. Khoze, M.~Spannowsky, and C.~Williams,
  ``Constraining dark sectors at colliders: Beyond the effective theory
  approach'',} \textit{ Phys. Rev. D} \textbf{ 91} (2015) 055009,
  \href{http://dx.doi.org/10.1103/PhysRevD.91.055009}{\doi{10.1103/PhysRevD.91.055009}},
\href{http://www.arXiv.org/abs/1411.0535}{\texttt{arXiv:1411.0535}}.

\bibitem{Haisch:2015ioa}
\hrefCMSnoop {}{U.~Haisch and E.~Re, ``{Simplified dark matter top-quark
  interactions at the {LHC}}'',} \textit{ JHEP} \textbf{ 06} (2015) 078,
  \href{http://dx.doi.org/10.1007/JHEP06(2015)078}{\doi{10.1007/JHEP06(2015)078}},
\href{http://www.arXiv.org/abs/1503.00691}{\texttt{arXiv:1503.00691}}.

\bibitem{Boveia:2016mrp}
\hrefCMSnoop {}{G.~Busoni {et~al.}, ``{Recommendations on presenting LHC
  searches for missing transverse energy signals using simplified $s$-channel
  models of dark matter}'',} (2016).
\href{http://www.arXiv.org/abs/1603.04156}{\texttt{arXiv:1603.04156}}.

\bibitem{CMS11}
\hrefCMSnoop {}{{CMS Collaboration}, ``Search for narrow resonances in dijet
  final states at $\sqrt{s}=$ 8 {TeV} with the novel {CMS} technique of data
  scouting'',} \textit{ Phys. Rev. Lett.} \textbf{ 117} (2016) 031802,
  \href{http://dx.doi.org/10.1103/PhysRevLett.117.031802}{\doi{10.1103/PhysRevLett.117.031802}},
\href{http://www.arXiv.org/abs/1604.08907}{\texttt{arXiv:1604.08907}}.

\bibitem{Sirunyan:2016iap}
\hrefCMSnoop {}{{CMS Collaboration}, ``{Search for dijet resonances in
  proton-proton collisions at $\sqrt{s}$ = 13 TeV and constraints on dark
  matter and other models}'',} \textit{ Phys. Lett. B} \textbf{ 769} (2017)
  520,
  \href{http://dx.doi.org/10.1016/j.physletb.2017.02.012}{\doi{10.1016/j.physletb.2017.02.012}},
  \href{http://www.arXiv.org/abs/1611.03568}{\texttt{arXiv:1611.03568}}.
[Erratum: \DOI{10.1016/j.physletb.2017.02.012}].

\bibitem{Sirunyan:2018xlo}
\hrefCMSnoop {}{{CMS Collaboration}, ``{Search for narrow and broad dijet
  resonances in proton-proton collisions at $\sqrt{s}=13$ TeV and constraints
  on dark matter mediators and other new particles}'',} \textit{ JHEP} \textbf{
  08} (2018) 130,
  \href{http://dx.doi.org/10.1007/JHEP08(2018)130}{\doi{10.1007/JHEP08(2018)130}},
\href{http://www.arXiv.org/abs/1806.00843}{\texttt{arXiv:1806.00843}}.

\bibitem{Aaboud:2018fzt}
\hrefCMSnoop {}{{ATLAS Collaboration}, ``{Search for low-mass dijet resonances
  using trigger-level jets with the ATLAS detector in $pp$ collisions at
  $\sqrt{s}=13$ TeV}'',} \textit{ Phys. Rev. Lett.} \textbf{ 121} (2018)
  081801,
  \href{http://dx.doi.org/10.1103/PhysRevLett.121.081801}{\doi{10.1103/PhysRevLett.121.081801}},
\href{http://www.arXiv.org/abs/1804.03496}{\texttt{arXiv:1804.03496}}.

\bibitem{Sirunyan:2018pas}
\hrefCMSnoop {}{{CMS Collaboration}, ``{Search for narrow resonances in the
  b-tagged dijet mass spectrum in proton-proton collisions at $\sqrt{s} =$ 8
  TeV}'',} \textit{ Phys. Rev. Lett.} \textbf{ 120} (2018) 201801,
  \href{http://dx.doi.org/10.1103/PhysRevLett.120.201801}{\doi{10.1103/PhysRevLett.120.201801}},
\href{http://www.arXiv.org/abs/1802.06149}{\texttt{arXiv:1802.06149}}.

\bibitem{Aaboud:2018tqo}
\hrefCMSnoop {}{{ATLAS Collaboration}, ``{Search for resonances in the mass
  distribution of jet pairs with one or two jets identified as $b$-jets in
  proton-proton collisions at $\sqrt{s}=13$ TeV with the ATLAS detector}'',}
  \textit{ Phys. Rev. D} \textbf{ 98} (2018) 032016,
  \href{http://dx.doi.org/10.1103/PhysRevD.98.032016}{\doi{10.1103/PhysRevD.98.032016}},
\href{http://www.arXiv.org/abs/1805.09299}{\texttt{arXiv:1805.09299}}.

\bibitem{Sirunyan:2017dnz}
\hrefCMSnoop {}{{CMS Collaboration}, ``Search for low mass vector resonances
  decaying to quark-antiquark pairs in proton-proton collisions at
  $\sqrt{s}=13\text{ }\text{ }\mathrm{TeV}$'',} \textit{ Phys. Rev. Lett.}
  \textbf{ 119} (2017) 111802,
  \href{http://dx.doi.org/10.1103/PhysRevLett.119.111802}{\doi{10.1103/PhysRevLett.119.111802}},
\href{http://www.arXiv.org/abs/1705.10532}{\texttt{arXiv:1705.10532}}.

\bibitem{Sirunyan:2017nvi}
\hrefCMSnoop {}{{CMS Collaboration}, ``Search for low mass vector resonances
  decaying into quark-antiquark pairs in proton-proton collisions at
  $\sqrt{s}=13\text{ }\text{ }\mathrm{TeV}$'',} \textit{ JHEP} \textbf{ 01}
  (2018) 097,
  \href{http://dx.doi.org/10.1007/JHEP01(2018)097}{\doi{10.1007/JHEP01(2018)097}},
\href{http://www.arXiv.org/abs/1710.00159}{\texttt{arXiv:1710.00159}}.

\bibitem{Aaboud:2018zba}
\hrefCMSnoop {}{{ATLAS Collaboration}, ``Search for light resonances decaying
  to boosted quark pairs and produced in association with a photon or a jet in
  proton-proton collisions at $\sqrt{s}=13$ {TeV} with the {ATLAS} detector'',}
  (2018).
  \href{http://www.arXiv.org/abs/1801.08769}{\texttt{arXiv:1801.08769}}.
Submitted to \textit{Phys. Lett. B}.

\bibitem{Sirunyan:2017dgc}
\hrefCMSnoop {}{{CMS Collaboration}, ``Inclusive search for a highly boosted
  {H}iggs boson decaying to a bottom quark-antiquark pair'',} \textit{ Phys.
  Rev. Lett.} \textbf{ 120} (2018) 071802,
  \href{http://dx.doi.org/10.1103/PhysRevLett.120.071802}{\doi{10.1103/PhysRevLett.120.071802}},
\href{http://www.arXiv.org/abs/1709.05543}{\texttt{arXiv:1709.05543}}.

\bibitem{Liu:2017xmc}
\hrefCMSnoop {}{D.~Liu, J.~Liu, C.~E.~M. Wagner, and X.-P. Wang, ``Bottom-quark
  forward-backward asymmetry, dark matter and the {LHC}'',} \textit{ Phys. Rev.
  D} \textbf{ 97} (2018) 055021,
  \href{http://dx.doi.org/10.1103/PhysRevD.97.055021}{\doi{10.1103/PhysRevD.97.055021}},
\href{http://www.arXiv.org/abs/1712.05802}{\texttt{arXiv:1712.05802}}.

\bibitem{Cacciari:2008gp}
\hrefCMSnoop {}{M.~Cacciari, G.~P. Salam, and G.~Soyez, ``The anti-$\kt$ jet
  clustering algorithm'',} \textit{ JHEP} \textbf{ 04} (2008) 063,
  \href{http://dx.doi.org/10.1088/1126-6708/2008/04/063}{\doi{10.1088/1126-6708/2008/04/063}},
\href{http://www.arXiv.org/abs/0802.1189}{\texttt{arXiv:0802.1189}}.

\bibitem{Dokshitzer:1997in}
\hrefCMSnoop {}{Y.~L. Dokshitzer, G.~D. Leder, S.~Moretti, and B.~R. Webber,
  ``Better jet clustering algorithms'',} \textit{ JHEP} \textbf{ 08} (1997)
  001,
  \href{http://dx.doi.org/10.1088/1126-6708/1997/08/001}{\doi{10.1088/1126-6708/1997/08/001}},
\href{http://www.arXiv.org/abs/hep-ph/9707323}{\texttt{arXiv:hep-ph/9707323}}.

\bibitem{Wobisch:1998wt}
\href {https://inspirehep.net/record/484872}{M.~Wobisch and T.~Wengler,
  ``Hadronization corrections to jet cross-sections in deep inelastic
  scattering'',} in \textit{ {Proceedings of the Workshop on Monte Carlo
  Generators for HERA Physics, Hamburg, Germany}}, p.~270.
\newblock 1998.
\newblock
\href{http://www.arXiv.org/abs/hep-ph/9907280}{\texttt{arXiv:hep-ph/9907280}}.
\newblock

\bibitem{Butterworth:2008iy}
\hrefCMSnoop {}{J.~M. Butterworth, A.~R. Davison, M.~Rubin, and G.~P. Salam,
  ``Jet substructure as a new {H}iggs search channel at the {LHC}'',} \textit{
  Phys. Rev. Lett.} \textbf{ 100} (2008) 242001,
  \href{http://dx.doi.org/10.1103/PhysRevLett.100.242001}{\doi{10.1103/PhysRevLett.100.242001}},
\href{http://www.arXiv.org/abs/0802.2470}{\texttt{arXiv:0802.2470}}.

\bibitem{Moult:2016cvt}
\hrefCMSnoop {}{I.~Moult, L.~Necib, and J.~Thaler, ``New angles on energy
  correlation functions'',} \textit{ JHEP} \textbf{ 12} (2016) 153,
  \href{http://dx.doi.org/10.1007/JHEP12(2016)153}{\doi{10.1007/JHEP12(2016)153}},
\href{http://www.arXiv.org/abs/1609.07483}{\texttt{arXiv:1609.07483}}.

\bibitem{Sirunyan:2017ezt}
\hrefCMSnoop {}{{CMS Collaboration}, ``Identification of heavy-flavour jets
  with the {CMS} detector in pp collisions at 13 {TeV}'',} \textit{ JINST}
  \textbf{ 13} (2018) P05011,
  \href{http://dx.doi.org/10.1088/1748-0221/13/05/P05011}{\doi{10.1088/1748-0221/13/05/P05011}},
\href{http://www.arXiv.org/abs/1712.07158}{\texttt{arXiv:1712.07158}}.

\bibitem{Khachatryan:2016bia}
\hrefCMSnoop {}{{CMS Collaboration}, ``The {CMS} trigger system'',} \textit{
  JINST} \textbf{ 12} (2017) P01020,
  \href{http://dx.doi.org/10.1088/1748-0221/12/01/P01020}{\doi{10.1088/1748-0221/12/01/P01020}},
\href{http://www.arXiv.org/abs/1609.02366}{\texttt{arXiv:1609.02366}}.

\bibitem{Chatrchyan:2008zzk}
\hrefCMSnoop {}{{CMS Collaboration}, ``The {CMS} experiment at the {CERN}
  {LHC}'',} \textit{ JINST} \textbf{ 3} (2008) S08004,
\href{http://dx.doi.org/10.1088/1748-0221/3/08/S08004}{\doi{10.1088/1748-0221/3/08/S08004}}.

\bibitem{GEANT4}
\hrefCMSnoop {}{{GEANT4} Collaboration, ``{\GEANTfour}---a simulation
  toolkit'',} \textit{ Nucl. Instrum. Meth. A} \textbf{ 506} (2003) 250,
\href{http://dx.doi.org/10.1016/S0168-9002(03)01368-8}{\doi{10.1016/S0168-9002(03)01368-8}}.

\bibitem{Alwall:2014hca}
J.~Alwall\hrefCMSnoop {}{ {et~al.}, ``The automated computation of tree-level
  and next-to-leading order differential cross sections, and their matching to
  parton shower simulations'',} \textit{ JHEP} \textbf{ 07} (2014) 079,
  \href{http://dx.doi.org/10.1007/JHEP07(2014)079}{\doi{10.1007/JHEP07(2014)079}},
  \href{http://www.arXiv.org/abs/1405.0301}{\texttt{arXiv:1405.0301}}.

\bibitem{Mattelaer:2015haa}
\hrefCMSnoop {}{O.~Mattelaer and E.~Vryonidou, ``{Dark matter production
  through loop-induced processes at the LHC: the $s$-channel mediator case}'',}
  \textit{ Eur. Phys. J. C} \textbf{ 75} (2015) 436,
  \href{http://dx.doi.org/10.1140/epjc/s10052-015-3665-5}{\doi{10.1140/epjc/s10052-015-3665-5}},
\href{http://www.arXiv.org/abs/1508.00564}{\texttt{arXiv:1508.00564}}.

\bibitem{Jones:2018hbb}
\hrefCMSnoop {}{S.~P. Jones, M.~Kerner, and G.~Luisoni,
  ``{Next-to-Leading-Order QCD Corrections to Higgs Boson Plus Jet Production
  with Full Top-Quark Mass Dependence}'',} \textit{ Phys. Rev. Lett.} \textbf{
  120} (2018) 162001,
  \href{http://dx.doi.org/10.1103/PhysRevLett.120.162001}{\doi{10.1103/PhysRevLett.120.162001}},
\href{http://www.arXiv.org/abs/1802.00349}{\texttt{arXiv:1802.00349}}.

\bibitem{Alwall:2007fs}
\hrefCMSnoop {}{J.~Alwall {et~al.}, ``Comparative study of various algorithms
  for the merging of parton showers and matrix elements in hadronic
  collisions'',} \textit{ Eur. Phys. J. C} \textbf{ 53} (2008) 473,
  \href{http://dx.doi.org/10.1140/epjc/s10052-007-0490-5}{\doi{10.1140/epjc/s10052-007-0490-5}},
\href{http://www.arXiv.org/abs/0706.2569}{\texttt{arXiv:0706.2569}}.

\bibitem{Nason:2004rx}
\hrefCMSnoop {}{P.~Nason, ``A new method for combining {NLO QCD} with shower
  {Monte Carlo} algorithms'',} \textit{ JHEP} \textbf{ 11} (2004) 040,
  \href{http://dx.doi.org/10.1088/1126-6708/2004/11/040}{\doi{10.1088/1126-6708/2004/11/040}},
\href{http://www.arXiv.org/abs/hep-ph/0409146}{\texttt{arXiv:hep-ph/0409146}}.

\bibitem{Frixione:2007vw}
\hrefCMSnoop {}{S.~Frixione, P.~Nason, and C.~Oleari, ``Matching {NLO} {QCD}
  computations with parton shower simulations: the {POWHEG} method'',} \textit{
  JHEP} \textbf{ 11} (2007) 070,
  \href{http://dx.doi.org/10.1088/1126-6708/2007/11/070}{\doi{10.1088/1126-6708/2007/11/070}},
\href{http://www.arXiv.org/abs/0709.2092}{\texttt{arXiv:0709.2092}}.

\bibitem{Alioli:2010xd}
\hrefCMSnoop {}{S.~Alioli, P.~Nason, C.~Oleari, and E.~Re, ``A general
  framework for implementing {NLO} calculations in shower monte carlo programs:
  the {POWHEG} {BOX}'',} \textit{ JHEP} \textbf{ 06} (2010) 043,
  \href{http://dx.doi.org/10.1007/JHEP06(2010)043}{\doi{10.1007/JHEP06(2010)043}},
\href{http://www.arXiv.org/abs/1002.2581}{\texttt{arXiv:1002.2581}}.

\bibitem{Luisoni:2013kna}
\hrefCMSnoop {}{G.~Luisoni, P.~Nason, C.~Oleari, and F.~Tramontano,
  ``{$\PH\PW^{\pm}$/$\PH\PZ$ + 0 and 1 jet at {NLO} with the {POWHEG} BOX
  interfaced to GoSam and their merging within MiNLO}'',} \textit{ JHEP}
  \textbf{ 10} (2013) 083,
  \href{http://dx.doi.org/10.1007/JHEP10(2013)083}{\doi{10.1007/JHEP10(2013)083}},
\href{http://www.arXiv.org/abs/1306.2542}{\texttt{arXiv:1306.2542}}.

\bibitem{deFlorian:2012mx}
\hrefCMSnoop {}{D.~de~Florian, G.~Ferrera, M.~Grazzini, and D.~Tommasini,
  ``{Higgs boson production at the LHC: transverse momentum resummation effects
  in the $\PH\to 2\gamma$, $\PH\to\PW\PW\to \ell\nu\ell\nu$ and
  $\PH\to\PZ\PZ\to 4\ell$ decay modes}'',} \textit{ JHEP} \textbf{ 06} (2012)
  132,
  \href{http://dx.doi.org/10.1007/JHEP06(2012)132}{\doi{10.1007/JHEP06(2012)132}},
\href{http://www.arXiv.org/abs/1203.6321}{\texttt{arXiv:1203.6321}}.

\bibitem{Grazzini:2013mca}
\hrefCMSnoop {}{M.~Grazzini and H.~Sargsyan, ``{Heavy-quark mass effects in
  Higgs boson production at the LHC}'',} \textit{ JHEP} \textbf{ 09} (2013)
  129,
  \href{http://dx.doi.org/10.1007/JHEP09(2013)129}{\doi{10.1007/JHEP09(2013)129}},
\href{http://www.arXiv.org/abs/1306.4581}{\texttt{arXiv:1306.4581}}.

\bibitem{Bagnaschi2012}
\hrefCMSnoop {}{E.~Bagnaschi, G.~Degrassi, P.~Slavich, and A.~Vicini, ``{Higgs
  production via gluon fusion in the {POWHEG} approach in the {SM} and in the
  {MSSM}}'',} \textit{ JHEP} \textbf{ 02} (2012) 088,
  \href{http://dx.doi.org/10.1007/JHEP02(2012)088}{\doi{10.1007/JHEP02(2012)088}},
\href{http://www.arXiv.org/abs/1111.2854}{\texttt{arXiv:1111.2854}}.

\bibitem{Bagnaschi:2015qta}
\hrefCMSnoop {}{E.~Bagnaschi and A.~Vicini, ``{The Higgs transverse momentum
  distribution in gluon fusion as a multiscale problem}'',} \textit{ JHEP}
  \textbf{ 01} (2016) 056,
  \href{http://dx.doi.org/10.1007/JHEP01(2016)056}{\doi{10.1007/JHEP01(2016)056}},
\href{http://www.arXiv.org/abs/1505.00735}{\texttt{arXiv:1505.00735}}.

\bibitem{Sjostrand:2014zea}
T.~Sj{\"o}strand\hrefCMSnoop {}{ {et~al.}, ``An introduction to {PYTHIA}
  8.2'',} \textit{ Comput. Phys. Commun.} \textbf{ 191} (2015) 159,
  \href{http://dx.doi.org/10.1016/j.cpc.2015.01.024}{\doi{10.1016/j.cpc.2015.01.024}},
\href{http://www.arXiv.org/abs/1410.3012}{\texttt{arXiv:1410.3012}}.

\bibitem{Khachatryan:2015pea}
\hrefCMSnoop {}{{CMS Collaboration}, ``Event generator tunes obtained from
  underlying event and multiparton scattering measurements'',} \textit{ Eur.
  Phys. J. C} \textbf{ 76} (2016) 155,
  \href{http://dx.doi.org/10.1140/epjc/s10052-016-3988-x}{\doi{10.1140/epjc/s10052-016-3988-x}},
\href{http://www.arXiv.org/abs/1512.00815}{\texttt{arXiv:1512.00815}}.

\bibitem{Campbell:2010ff}
\hrefCMSnoop {}{J.~M. Campbell and R.~K. Ellis, ``{MCFM} for the {T}evatron and
  the {LHC}'',} \textit{ Nucl. Phys. Proc. Suppl.} \textbf{ 205-206} (2010) 10,
  \href{http://dx.doi.org/10.1016/j.nuclphysbps.2010.08.011}{\doi{10.1016/j.nuclphysbps.2010.08.011}},
\href{http://www.arXiv.org/abs/1007.3492}{\texttt{arXiv:1007.3492}}.

\bibitem{Czakon:2013goa}
\hrefCMSnoop {}{M.~Czakon, P.~Fiedler, and A.~Mitov, ``Total top quark
  pair-production cross section at hadron colliders through
  $\mathcal{O}(\alpha_s^4)$'',} \textit{ Phys. Rev. Lett.} \textbf{ 110} (2013)
  252004,
  \href{http://dx.doi.org/10.1103/PhysRevLett.110.252004}{\doi{10.1103/PhysRevLett.110.252004}},
\href{http://www.arXiv.org/abs/1303.6254}{\texttt{arXiv:1303.6254}}.

\bibitem{Kallweit:2014xda}
S.~Kallweit\hrefCMSnoop {}{ {et~al.}, ``{NLO} electroweak automation and
  precise predictions for {$\PW$}+multijet production at the {LHC}'',} \textit{
  JHEP} \textbf{ 04} (2015) 012,
  \href{http://dx.doi.org/10.1007/JHEP04(2015)012}{\doi{10.1007/JHEP04(2015)012}},
\href{http://www.arXiv.org/abs/1412.5157}{\texttt{arXiv:1412.5157}}.

\bibitem{Kallweit:2015dum}
S.~Kallweit\hrefCMSnoop {}{ {et~al.}, ``{NLO} {QCD+EW} predictions for {V} +
  jets including off-shell vector-boson decays and multijet merging'',}
  \textit{ JHEP} \textbf{ 04} (2016) 021,
  \href{http://dx.doi.org/10.1007/JHEP04(2016)021}{\doi{10.1007/JHEP04(2016)021}},
\href{http://www.arXiv.org/abs/1511.08692}{\texttt{arXiv:1511.08692}}.

\bibitem{Kallweit:2015fta}
S.~Kallweit\href
  {https://inspirehep.net/record/1372103/files/arXiv:1505.05704.pdf}{ {et~al.},
  ``{NLO} {QCD+EW} automation and precise predictions for {V}+multijet
  production'',} in \textit{ {Proceedings, 50th Rencontres de Moriond, QCD and
  high energy interactions}}, p.~121.
\newblock 2015.
\newblock
\href{http://www.arXiv.org/abs/1505.05704}{\texttt{arXiv:1505.05704}}.
\newblock

\bibitem{Lindert:2017olm}
\hrefCMSnoop {}{J.~M. Lindert {et~al.}, ``{Precise predictions for $\PV$+jets
  dark matter backgrounds}'',} \textit{ Eur. Phys. J. C} \textbf{ 77} (2017)
  829,
  \href{http://dx.doi.org/10.1140/epjc/s10052-017-5389-1}{\doi{10.1140/epjc/s10052-017-5389-1}},
\href{http://www.arXiv.org/abs/1705.04664}{\texttt{arXiv:1705.04664}}.

\bibitem{Ball:2014uwa}
\hrefCMSnoop {}{{NNPDF} Collaboration, ``Parton distributions for the {LHC} run
  {II}'',} \textit{ JHEP} \textbf{ 04} (2015) 040,
  \href{http://dx.doi.org/10.1007/JHEP04(2015)040}{\doi{10.1007/JHEP04(2015)040}},
\href{http://www.arXiv.org/abs/1410.8849}{\texttt{arXiv:1410.8849}}.

\bibitem{Sirunyan:2017ulk}
\hrefCMSnoop {}{{CMS Collaboration}, ``Particle-flow reconstruction and global
  event description with the {CMS} detector'',} \textit{ JINST} \textbf{ 12}
  (2017) P10003,
  \href{http://dx.doi.org/10.1088/1748-0221/12/10/P10003}{\doi{10.1088/1748-0221/12/10/P10003}},
\href{http://www.arXiv.org/abs/1706.04965}{\texttt{arXiv:1706.04965}}.

\bibitem{Cacciari:2011ma}
\hrefCMSnoop {}{M.~Cacciari, G.~P. Salam, and G.~Soyez, ``Fastjet user
  manual'',} \textit{ Eur. Phys. J. C} \textbf{ 72} (2012) 1896,
  \href{http://dx.doi.org/10.1140/epjc/s10052-012-1896-2}{\doi{10.1140/epjc/s10052-012-1896-2}},
\href{http://www.arXiv.org/abs/1111.6097}{\texttt{arXiv:1111.6097}}.

\bibitem{Bertolini:2014bba}
\hrefCMSnoop {}{D.~Bertolini, P.~Harris, M.~Low, and N.~Tran, ``{P}ileup {P}er
  {P}article {I}dentification'',} \textit{ JHEP} \textbf{ 10} (2014) 059,
  \href{http://dx.doi.org/10.1007/JHEP10(2014)059}{\doi{10.1007/JHEP10(2014)059}},
\href{http://www.arXiv.org/abs/1407.6013}{\texttt{arXiv:1407.6013}}.

\bibitem{Khachatryan:2016kdb}
\hrefCMSnoop {}{{CMS Collaboration}, ``{Jet energy scale and resolution in the
  CMS experiment in pp collisions at 8 TeV}'',} \textit{ JINST} \textbf{ 12}
  (2017) P02014,
  \href{http://dx.doi.org/10.1088/1748-0221/12/02/P02014}{\doi{10.1088/1748-0221/12/02/P02014}},
\href{http://www.arXiv.org/abs/1607.03663}{\texttt{arXiv:1607.03663}}.

\bibitem{Krohn:2009th}
\hrefCMSnoop {}{D.~Krohn, J.~Thaler, and L.-T. Wang, ``Jet trimming'',}
  \textit{ JHEP} \textbf{ 02} (2010) 084,
  \href{http://dx.doi.org/10.1007/JHEP02(2010)084}{\doi{10.1007/JHEP02(2010)084}},
\href{http://www.arXiv.org/abs/0912.1342}{\texttt{arXiv:0912.1342}}.

\bibitem{Khachatryan:2015hwa}
\hrefCMSnoop {}{{CMS Collaboration}, ``Performance of electron reconstruction
  and selection with the {CMS} detector in proton-proton collisions at
  $\sqrt{s}=$ 8 {TeV}'',} \textit{ JINST} \textbf{ 10} (2015) P06005,
  \href{http://dx.doi.org/10.1088/1748-0221/10/06/P06005}{\doi{10.1088/1748-0221/10/06/P06005}},
\href{http://www.arXiv.org/abs/1502.02701}{\texttt{arXiv:1502.02701}}.

\bibitem{Sirunyan:2018fpa}
\hrefCMSnoop {}{{CMS Collaboration}, ``{Performance of the CMS muon detector
  and muon reconstruction with proton-proton collisions at $\sqrt{s}=$ 13
  TeV}'',} \textit{ JINST} \textbf{ 13} (2018) P06015,
  \href{http://dx.doi.org/10.1088/1748-0221/13/06/P06015}{\doi{10.1088/1748-0221/13/06/P06015}},
\href{http://www.arXiv.org/abs/1804.04528}{\texttt{arXiv:1804.04528}}.

\bibitem{Larkoski:2014wba}
\hrefCMSnoop {}{A.~J. Larkoski, S.~Marzani, G.~Soyez, and J.~Thaler, ``Soft
  drop'',} \textit{ JHEP} \textbf{ 05} (2014) 146,
  \href{http://dx.doi.org/10.1007/JHEP05(2014)146}{\doi{10.1007/JHEP05(2014)146}},
\href{http://www.arXiv.org/abs/1402.2657}{\texttt{arXiv:1402.2657}}.

\bibitem{Dasgupta:2013ihk}
\hrefCMSnoop {}{M.~Dasgupta, A.~Fregoso, S.~Marzani, and G.~P. Salam, ``Towards
  an understanding of jet substructure'',} \textit{ JHEP} \textbf{ 09} (2013)
  029,
  \href{http://dx.doi.org/10.1007/JHEP09(2013)029}{\doi{10.1007/JHEP09(2013)029}},
\href{http://www.arXiv.org/abs/1307.0007}{\texttt{arXiv:1307.0007}}.

\bibitem{CMS-PAS-JME-16-003}
\href {https://cds.cern.ch/record/2256875}{{CMS Collaboration}, ``{Jet
  algorithms performance in 13 TeV data}'',} CMS Physics Analysis Summary
  CMS-PAS-JME-16-003, CERN, Geneva, 2017.

\bibitem{Dolen:2016kst}
J.~Dolen\hrefCMSnoop {}{ {et~al.}, ``Thinking outside the {ROC}s: Designing
  decorrelated taggers ({DDT}) for jet substructure'',} \textit{ JHEP} \textbf{
  05} (2016) 156,
  \href{http://dx.doi.org/10.1007/JHEP05(2016)156}{\doi{10.1007/JHEP05(2016)156}},
\href{http://www.arXiv.org/abs/1603.00027}{\texttt{arXiv:1603.00027}}.

\bibitem{Larkoski:2013eya}
\hrefCMSnoop {}{A.~J. Larkoski, G.~P. Salam, and J.~Thaler, ``Energy
  correlation functions for jet substructure'',} \textit{ JHEP} \textbf{ 06}
  (2013) 108,
  \href{http://dx.doi.org/10.1007/JHEP06(2013)108}{\doi{10.1007/JHEP06(2013)108}},
\href{http://www.arXiv.org/abs/1305.0007}{\texttt{arXiv:1305.0007}}.

\bibitem{Thaler:2010tr}
\hrefCMSnoop {}{J.~Thaler and K.~Van~Tilburg, ``Identifying boosted objects
  with {N}-subjettiness'',} \textit{ JHEP} \textbf{ 03} (2011) 015,
  \href{http://dx.doi.org/10.1007/JHEP03(2011)015}{\doi{10.1007/JHEP03(2011)015}},
\href{http://www.arXiv.org/abs/1011.2268}{\texttt{arXiv:1011.2268}}.

\bibitem{ref:ftest}
\hrefCMSnoop {}{R.~A. Fisher, ``On the interpretation of $\chi^{2}$ from
  contingency tables, and the calculation of {P}'',} \textit{ J. Royal Stat.
  Soc.} \textbf{ 85} (1922) 87,
  \href{http://dx.doi.org/10.2307/2340521}{\doi{10.2307/2340521}}.

\bibitem{Sirunyan:2017jix}
\hrefCMSnoop {}{{CMS Collaboration}, ``{Search for new physics in final states
  with an energetic jet or a hadronically decaying $\PW$ or $\PZ$ boson and
  transverse momentum imbalance at $\sqrt{s} =$ 13 TeV}'',} \textit{ Phys. Rev.
  D} \textbf{ 97} (2018) 092005,
  \href{http://dx.doi.org/10.1103/PhysRevD.97.092005}{\doi{10.1103/PhysRevD.97.092005}},
\href{http://www.arXiv.org/abs/1712.02345}{\texttt{arXiv:1712.02345}}.

\bibitem{Denner:2009gj}
\hrefCMSnoop {}{A.~Denner, S.~Dittmaier, T.~Kasprzik, and A.~Muck,
  ``{Electroweak corrections to $\PW$+jet hadroproduction including leptonic
  $\PW$-boson decays}'',} \textit{ JHEP} \textbf{ 08} (2009) 075,
  \href{http://dx.doi.org/10.1088/1126-6708/2009/08/075}{\doi{10.1088/1126-6708/2009/08/075}},
\href{http://www.arXiv.org/abs/0906.1656}{\texttt{arXiv:0906.1656}}.

\bibitem{Denner:2011vu}
\hrefCMSnoop {}{A.~Denner, S.~Dittmaier, T.~Kasprzik, and A.~Muck,
  ``{Electroweak corrections to dilepton+jet production at hadron
  colliders}'',} \textit{ JHEP} \textbf{ 06} (2011) 069,
  \href{http://dx.doi.org/10.1007/JHEP06(2011)069}{\doi{10.1007/JHEP06(2011)069}},
\href{http://www.arXiv.org/abs/1103.0914}{\texttt{arXiv:1103.0914}}.

\bibitem{Denner:2012ts}
\hrefCMSnoop {}{A.~Denner, S.~Dittmaier, T.~Kasprzik, and A.~Maeck,
  ``{Electroweak corrections to monojet production at the LHC}'',} \textit{
  Eur. Phys. J. C} \textbf{ 73} (2013) 2297,
  \href{http://dx.doi.org/10.1140/epjc/s10052-013-2297-x}{\doi{10.1140/epjc/s10052-013-2297-x}},
\href{http://www.arXiv.org/abs/1211.5078}{\texttt{arXiv:1211.5078}}.

\bibitem{Kuhn:2005gv}
\hrefCMSnoop {}{J.~H. Kuhn, A.~Kulesza, S.~Pozzorini, and M.~Schulze,
  ``{Electroweak corrections to hadronic photon production at large transverse
  momenta}'',} \textit{ JHEP} \textbf{ 03} (2006) 059,
  \href{http://dx.doi.org/10.1088/1126-6708/2006/03/059}{\doi{10.1088/1126-6708/2006/03/059}},
\href{http://www.arXiv.org/abs/hep-ph/0508253}{\texttt{arXiv:hep-ph/0508253}}.

\bibitem{jec}
\hrefCMSnoop {}{{CMS Collaboration}, ``{Determination of jet energy calibration
  and transverse momentum resolution in CMS}'',} \textit{ JINST} \textbf{ 6}
  (2011) 11002,
  \href{http://dx.doi.org/10.1088/1748-0221/6/11/P11002}{\doi{10.1088/1748-0221/6/11/P11002}},
  \href{http://www.arXiv.org/abs/1107.4277}{\texttt{arXiv:1107.4277}}.

\bibitem{lumi}
\href {https://cds.cern.ch/record/2257069}{{CMS Collaboration}, ``{CMS}
  luminosity measurements for the 2016 data taking period'',} CMS Physics
  Analysis Summary CMS-PAS-LUM-17-001, CERN, 2017.

\bibitem{LHCCLs}
\href {http://cdsweb.cern.ch/record/1379837}{{ATLAS and CMS Collaborations},
  ``{Procedure for the LHC Higgs boson search combination in Summer 2011}'',}
  CMS Note/ATLAS Pub CMS-NOTE-2011-005, ATL-PHYS-PUB-2011-11, CERN, 2011.

\bibitem{CLS1}
\hrefCMSnoop {}{A.~L. Read, ``{Presentation of search results: the CL$_s$
  technique}'',} \textit{ J. Phys. G} \textbf{ 28} (2002) 2693,
  \href{http://dx.doi.org/10.1088/0954-3899/28/10/313}{\doi{10.1088/0954-3899/28/10/313}}.

\bibitem{CLS2}
\hrefCMSnoop {}{T.~Junk, ``{Confidence level computation for combining searches
  with small statistics}'',} \textit{ Nucl. Instrum. Meth. A} \textbf{ 434}
  (1999) 435,
  \href{http://dx.doi.org/10.1016/S0168-9002(99)00498-2}{\doi{10.1016/S0168-9002(99)00498-2}},
  \href{http://www.arXiv.org/abs/hep-ex/9902006}{\texttt{arXiv:hep-ex/9902006}}.

\bibitem{Khachatryan:2014jba}
\hrefCMSnoop {}{{CMS Collaboration}, ``{Precise determination of the mass of
  the Higgs boson and tests of compatibility of its couplings with the standard
  model predictions using proton collisions at 7 and 8 TeV}'',} \textit{ Eur.
  Phys. J. C} \textbf{ 75} (2015) 212,
  \href{http://dx.doi.org/10.1140/epjc/s10052-015-3351-7}{\doi{10.1140/epjc/s10052-015-3351-7}},
\href{http://www.arXiv.org/abs/1412.8662}{\texttt{arXiv:1412.8662}}.

\bibitem{Cowan:2010js}
\hrefCMSnoop {}{G.~Cowan, K.~Cranmer, E.~Gross, and O.~Vitells, ``Asymptotic
  formulae for likelihood-based tests of new physics'',} \textit{ Eur. Phys. J.
  C} \textbf{ 71} (2011) 1554,
  \href{http://dx.doi.org/10.1140/epjc/s10052-011-1554-0}{\doi{10.1140/epjc/s10052-011-1554-0}},
  \href{http://www.arXiv.org/abs/1007.1727}{\texttt{arXiv:1007.1727}}.
[Erratum: \DOI{10.1140/epjc/s10052-013-2501-z}].

\bibitem{pvalue}
\hrefCMSnoop {}{L.~Demortier, ``{P} values and nuisance parameters'',} in
  \textit{ Statistical issues for {LHC} physics. {Proceedings, Workshop,
  PHYSTAT-LHC, Geneva, Switzerland, June} 27-29, 2007}, p.~23.
\newblock 2008.
\newblock
\href{http://dx.doi.org/10.5170/CERN-2008-001}{\doi{10.5170/CERN-2008-001}}.

\end{thebibliography}\endgroup
\cleardoublepage \appendix\section{The CMS Collaboration \label{app:collab}}\begin{sloppypar}\hyphenpenalty=5000\widowpenalty=500\clubpenalty=5000\vskip\cmsinstskip
\textbf{Yerevan Physics Institute, Yerevan, Armenia}\\*[0pt]
A.M.~Sirunyan, A.~Tumasyan
\vskip\cmsinstskip
\textbf{Institut f\"{u}r Hochenergiephysik, Wien, Austria}\\*[0pt]
W.~Adam, F.~Ambrogi, E.~Asilar, T.~Bergauer, J.~Brandstetter, M.~Dragicevic, J.~Er\"{o}, A.~Escalante~Del~Valle, M.~Flechl, R.~Fr\"{u}hwirth\cmsAuthorMark{1}, V.M.~Ghete, J.~Hrubec, M.~Jeitler\cmsAuthorMark{1}, N.~Krammer, I.~Kr\"{a}tschmer, D.~Liko, T.~Madlener, I.~Mikulec, N.~Rad, H.~Rohringer, J.~Schieck\cmsAuthorMark{1}, R.~Sch\"{o}fbeck, M.~Spanring, D.~Spitzbart, A.~Taurok, W.~Waltenberger, J.~Wittmann, C.-E.~Wulz\cmsAuthorMark{1}, M.~Zarucki
% ...
\vskip\cmsinstskip
% ...
1:  Also at Vienna University of Technology, Vienna, Austria\\
% ...
\end{sloppypar}
\end{document}